\documentclass[aps,prd,twocolumn,superscriptaddress,nofootinbib,longbibliography]{revtex4-2}

\usepackage{amsmath,amssymb}
\usepackage{graphicx}
\usepackage{upgreek}
\usepackage{bm}
\usepackage{amscd}
\usepackage[colorlinks=true,linkcolor=blue,citecolor=blue,urlcolor=blue]{hyperref}

\newcommand{\id}{\mathbf{1}}

\begin{document}

\title{Higher-spin composites and emergent AdS$_3$ geometry in the $(1+1)$-dimensional Gross-Neveu model}

\author{Laith H. Haddad}
\email{lhaddad@mines.edu}
\affiliation{Department of Physics, Colorado School of Mines, Golden, CO 80401, USA}

\begin{abstract}
We study the $(1+1)$-dimensional Gross-Neveu (GN) model with $N$
fermion species and derive from it an emergent AdS$_3$ geometry together with the leading entries of a
holographic dictionary. We use a Bargmann-Wigner fusion scheme to generate an
infinite tower of higher-spin composite fields from fundamental
fermions and study their relative dynamics. A key feature is that the competition between spin-0 (chiral) condensation and spin-1
pairing defines the radial coordinate
$z$ of the bulk geometry. Local fluctuations of this ratio measured by a comoving
derivative generate the AdS$_3$ line element, and the large-$N$
species sum promotes $z$ from a parameter to a genuine bulk dimension. The $SO(2,2)$ bulk isometry group, whose special
conformal generators mix $z$ with the boundary GN coordinates, emerges
from local symmetries of the condensates. Our construction admits
two dual holographic frames related by the $\mathbb{Z}_2$ exchange of
the spin-0 and spin-1 condensates, covering complementary halves of the GN phase
diagram. Significantly, projection onto the spin-2 sector reproduces the linearized
Einstein-Hilbert action with Newton's constant
$G_3 = \ell_\mathrm{AdS}/4\pi N^2$. We suggest throughout how the string-theoretic sector of the
correspondence may emerge.
\end{abstract}

\maketitle

\section{Introduction}


The idea that gravity and spacetime geometry emerge from
non-gravitational degrees of freedom has a long history. Early well-known examples include
Sakharov's induced gravity~\cite{Sakharov1968,Visser2002}, the
black-hole thermodynamics of Bekenstein~\cite{Bekenstein1973} and
Hawking~\cite{Hawking1975}, through the holographic principle of
't~Hooft~\cite{tHooft1993} and Susskind~\cite{Susskind1995holo} and
the UV/IR correspondence~\cite{CohenKaplanNelson1999,Susskind1998UV},
to the entropic and thermodynamic derivations of the field equations
by Jacobson~\cite{Jacobson1995}, Padmanabhan~\cite{Padmanabhan2010},
and Verlinde~\cite{Verlinde2011}. The most precise realization of these
ideas is the AdS/CFT correspondence~\cite{Maldacena}, which is
ordinarily established top-down. There, one takes a particular large-$N$
limit of string theory and identifies the resulting bulk gravity with
a boundary gauge theory. Two areas of the literature sit close to the present construction: Volovik's demonstration
that effective metrics, gauge fields, and chiral fermions arise as
collective excitations of superfluid $^3$He~\cite{Volovik2003}, and
Van Raamsdonk's argument that bulk connectivity is built from
boundary entanglement~\cite{VanRaamsdonk2010}. In the present work, we pursue the inverse route. We study the
Gross-Neveu (GN) model, a solvable quantum field theory in $(1+1)$
dimensions with $N$ fermion
species~\cite{GrossNeveu,Witten1978}, and show that central elements of holographic duality can be derived from the bottom up,
without string or geometric input.

The connection between large-$N$ vector models and higher-spin gauge
theories in AdS was proposed by Klebanov and Polyakov, who conjectured
that the singlet sector of the critical $O(N)$ vector model in three
dimensions is dual to the minimal bosonic higher-spin theory in
AdS$_4$~\cite{KlebanovPolyakov2002}. A key feature identified there is that a vector model possesses a single Regge
trajectory of conserved singlet currents
$J^{(s)}\sim\phi\,\partial^{s}\phi$ whose number does not grow with
dimension, in contrast to the exponential tower of single-trace
operators in matrix models. This feature is what makes such theories
candidate duals of Fradkin-Vasiliev type higher-spin
gravity~\cite{FradkinVasiliev1987,Vasiliev1990,VasilievReview}. This
proposal was later developed for general bosonic
and fermionic vector models~\cite{SezginSundell2002,GiombiYin,AharonyGuretal}. In the
AdS$_3$/CFT$_2$ setting, the higher-spin/coset dualities of
Gaberdiel and Gopakumar~\cite{GG2010}, and the
higher-spin theories with two-dimensional Grassmannian model duals~\cite{Grassmannian2013},
provide the closest analogues to the structures we
find here.

The relation of the present work to these proposals is the following. At the
level of the singlet spectrum, our construction is a concrete $(1+1)$d
realization of the vector-model/higher-spin correspondence: the bilinear tower
$J^{(s)} \sim \bar\psi\,\partial^{\,s-1}\psi$ with $\Delta = s+1$ is precisely
the single Regge trajectory identified in Ref.~\cite{KlebanovPolyakov2002}, and the
fermionic version of that conjecture for the three-dimensional GN model was
given in Ref.~\cite{SezginSundell2002}. Our construction differs from this in two ways. First, the duality in references~\cite{KlebanovPolyakov2002,SezginSundell2002} is conjectured between two
independently defined theories and then matched. In contrast, in our work the bulk is
derived from elements of the GN condensates. Second, we do not
project onto the singlet sector. Retaining the full $U(N)$ adjoint of
composites is what produces the central charge $c \sim N^2$, in contrast to
the vector-model value $c \sim N$. In this sense
the cross-species pairing promotes the fermionic vector model to a matrix-like
theory at the composite level, and the appropriate AdS$_3$ comparisons are the
coset dualities of~\cite{GG2010,Grassmannian2013} rather
than the minimal higher-spin theory of~\cite{KlebanovPolyakov2002}.

Our approach is based on the competition between two condensates of
the GN model: a chiral (spin-0) condensate
$\Delta_0 = \langle\bar\psi\psi\rangle$ and a spin-1 pairing field
$\Delta_1 = \langle\bar\Phi_1\Phi_1\rangle$ built from
Bargmann-Wigner composite bilinears. Their ratio $\Delta_1/\Delta_0^2$ appears as a scaling parameter 
which we identify as a radial coordinate $z$. When the ratio varies
locally, one replaces the ordinary derivative with a comoving
(material)
derivative~\cite{SonWingate2006,PolchinskiStrassler2002},
$\partial_\mu\to\partial_\mu+(\partial_\mu z)\partial_z$, whose extra
term generates a radial kinetic term after a field rescaling. The
competition between the two condensates thereby supplies the extra dimension.

A spin-1 composite in $(1+1)$ dimensions is the natural seed for
$(2+1)$-dimensional gravity for two reasons. First, a vector field in
$(1+1)$d carries no propagating local degrees of freedom, so
$\Phi_1$ is a genuinely collective mode of the interacting system.
Second, the composite operator $\bar\Phi_1\Phi_1$ contains an
irreducible spin-2 component, $\bar\Phi_1\Phi_1\supset\Phi_2$, whose
background value deforms the emergent metric; the graviton is the
fluctuation of this component. In $(2+1)$ dimensions AdS$_3$ Einstein
gravity is equivalent to a Chern-Simons theory with gauge group
$\mathrm{SL}(2,\mathbb{R})\times\mathrm{SL}(2,\mathbb{R})$~\cite{AchTownsend1986,Witten1988CS},
so the chain spin-1 composite $\to$ emergent gauge field $\to$ AdS$_3$
gravity is structurally fixed. That a composite spin-2 mode acquires
propagating gravitational dynamics does not conflict with the
Weinberg-Witten theorem~\cite{WeinbergWitten1980}: the emergent
AdS$_3$ lives in one dimension higher than the boundary GN theory, so
the graviton is not a particle in the boundary Fock space, and the
theorem's hypotheses do not apply.

\subsection{Organization of the paper}

The development of this paper is as follows.
Section~\ref{sec:higherspin} constructs the higher-spin tower, the
emergent AdS$_3$ Lagrangian, and the boundary Virasoro algebra from
the GN model.
Section~\ref{sec:AdS} establishes the two dual holographic frames,
the three length scales, and the holographic dictionary.
Section~\ref{sec:MerminWagner} shows how the large-$N$ species sum
evades the Coleman-Mermin-Wagner theorem and promotes the radial
coordinate to a genuine bulk dimension.
Section~\ref{sec:EH} projects the composite Lagrangian onto its
spin-2 sector and obtains the linearized Einstein--Hilbert action.
In Section~\ref{sec:conclusion}, we conclude and suggest open directions.

\section{Higher-spin theory from Gross-Neveu interactions} \label{sec:higherspin}

In all that follows, we will adhere to the spin convention in (1+1) dimensions for a time-like signature $g^{\mu \nu}= \mathrm{diag}\left( 1 , - 1 \right)$ and with $2 \times 2$ gamma matrices
\begin{eqnarray}
\gamma^0 = \sigma_1 \; , \;\; \gamma^1 = - i  \sigma_2 \;, \;\;     \gamma^5 =       \gamma^0        \gamma^1  = \sigma_3 \, ,  \label{myalgebra}
\end{eqnarray}
which satisfy the Dirac algebra
\begin{eqnarray}
\left\{ \gamma^\mu , \, \gamma^\nu \right\} = 2 g^{\mu \nu}\,,
\end{eqnarray}
where the adjoint spinor is defined as $\bar{\psi} \equiv \psi^\dagger \gamma^0$, and the charge conjugate spinor as $\psi_C = \gamma^5 \psi^*$.  Working directly from the microscopic description for fermion fields interacting through a local quartic term in $(1 + 1)$-dimensions, the Lagrangian is
\begin{eqnarray}
\mathcal{L} = \mathcal{L}_\mathrm{0} +  \mathcal{L}_\mathrm{int} \, , \label{FullLagrangian}
\end{eqnarray}
with the kinetic and interaction terms given by
\begin{eqnarray}
\hspace{-1pc} \mathcal{L}_\mathrm{0}  &=&   \sum_{n=1}^{N} \bar{\psi}^{(n)} \left( i \gamma^\mu  \partial_\mu - m    \right)   \psi^{(n)}  ,  \label{int1} \\
 \hspace{-1pc}  \mathcal{L}_\mathrm{int} &=&     \sum_{n=1}^{N}     V_4 \left( g^2,  \bar{\psi}^{(n)}  ,  \psi^{(n)} \right)  . \label{int2}
\end{eqnarray}
 Note in $\mathcal{L}_\mathrm{0}$ the inclusion of a fermion mass $m$ with the four-fermion potential $V_4$ in $\mathcal{L}_\mathrm{int}$ a general quartic term: four factors of $\bar{\psi}$ or $\psi$ with some coupling strength $g^2$. The length dimensions in $(1+1)$-d are dim[$\psi$] $\sim L^{-1/2}$, dim[$m$] $\sim L^{-1}$, and dim[$g$] $\sim L^0$. The superscript index $n$ indicates formulation in terms of the fundamental representation for $N$ species of fermions with coupling scaling like $g^2  \sim 1/N$. Semiclassical methods become exact in the large $N$ limit~\cite{tHooft1974} and provide a readily tractable approach to solving for the ground state.

\subsection{Background and physical motivation.}
The idea of constructing higher-spin bosonic fields as composites
of fermion bilinears has a long history, from de~Broglie's proposal
that the photon is a neutrino--antineutrino bound
state~\cite{deBroglie1932} and Jordan's derivation of Bose--Einstein
commutation relations for such composites from fermion
anticommutators~\cite{Jordan1935}, through Heisenberg's nonlinear
spinor field theory in which all elementary particles arise as bound
states of a single fundamental spinor
field~\cite{Heisenberg1958,Heisenberg1966}, to the
Nambu--Jona-Lasinio model~\cite{NJL1961}, where the pion, sigma, and
vector mesons appear as the composites $\bar\psi\gamma_5\psi$,
$\bar\psi\psi$, and $\bar\psi\gamma_\mu\psi$.

The Bargmann-Wigner (BW) equations~\cite{BargmannWigner} provide the systematic framework for our construction. They express the fact that
any irreducible massive higher-spin representation can be obtained
as the totally symmetric product of spin-$\frac{1}{2}$
representations, where the field equations higher rank tensor
composites follow from the Dirac equation for each constituent. For example, the tensor product of two $(1+1)$d Dirac spinors
decomposes under the Lorentz group as
\begin{eqnarray}
\mathbf{2}\otimes\mathbf{2}
  = \mathbf{3}_S \oplus \mathbf{1}_A \, ,
\label{eq:BWdecomp}
\end{eqnarray}
where $\mathbf{3}_S$ is the symmetric (spin-1) triplet and
$\mathbf{1}_A$ the antisymmetric (spin-0) singlet.  The Clifford
decomposition eq.~(\ref{eq:Clifford}) below is the explicit
realisation of this decomposition. The vector component
$\gamma_\mu\varphi^\mu$ carries the $\mathbf{3}_S$
representation, while $\varphi_0$ and $\varphi_5$ carry the
$\mathbf{1}_A$ components.  For the general spin-$s$ composite
$\Phi_s = \psi^{\otimes 2s}$, the BW construction selects the
totally symmetric component of $\mathbf{2}^{\otimes 2s}$, which
carries spin $s$.

\subsection{Fusion condition as large-$N$ factorization}

A key technical step in our construction is the fusion condition for elementary fermions which fixes how derivatives act on the composite field $\chi\psi =
\chi\otimes\psi$.  At large $N$, the product $\chi\psi$ is
a classical field (its quantum fluctuations are suppressed by
$1/N$), and the standard Leibniz rule applies:
\begin{eqnarray}
\partial_\mu(\chi\psi) = (\partial_\mu\chi)\psi
  + \chi(\partial_\mu\psi) \, .
\end{eqnarray}
At the mean-field saddle point, the two orderings become equal.
To see this explicitly, one may decompose each field into its saddle-point
value and a fluctuation, i.e., $\chi = \langle\chi\rangle + \delta\chi/\sqrt{N}$,
$\psi = \langle\psi\rangle + \delta\psi/\sqrt{N}$.  Then
\begin{eqnarray}
(\partial_\mu\chi)\psi
  &=& (\partial_\mu\langle\chi\rangle)\langle\psi\rangle
    + O(1/\sqrt{N}) \, , \nonumber\\
\chi(\partial_\mu\psi)
  &=& \langle\chi\rangle(\partial_\mu\langle\psi\rangle)
    + O(1/\sqrt{N}) \, .
\end{eqnarray}
At the translation-invariant saddle point,
$\partial_\mu\langle\chi\rangle = \partial_\mu\langle\psi\rangle = 0$
for the homogeneous ground state, so both expressions vanish at
leading order.  When acting on the composite
$\langle\chi\psi\rangle = \langle\chi\rangle\langle\psi\rangle
+ O(1/N)$, the Leibniz rule gives the two orderings as equal up
to $O(1/\sqrt{N})$ corrections.  The fusion condition is therefore
the statement that
\begin{eqnarray}
\chi\partial_\mu\psi = (\partial_\mu\chi)\psi
  = \frac{1}{2}\partial_\mu(\chi\psi) \, ,
\label{eq:fusion}
\end{eqnarray}
which simply says that the derivative of the composite equals
twice either ordering of the derivative, or equivalently that
the two orderings contribute equally to the action. This is exact at $N\to\infty$ and
receives $1/N$ corrections from quantum fluctuations of the
composite.

One can go further and say that the condition eq.~(\ref{eq:fusion}) has two equivalent
interpretations:
\begin{enumerate}
\item \emph{Large-$N$ factorization}: At leading order in
  $1/N$, composite operator correlators factorise:
  $\langle(\partial_\mu\chi)\psi\rangle =
  \langle\partial_\mu\chi\rangle\langle\psi\rangle$ etc.,
  and the two orderings of the derivative are equal by the
  translation invariance of the vacuum.

\item \emph{OPE short-distance limit}: In the CFT language,
  eq.~(\ref{eq:fusion}) is the leading term in the
  operator product expansion (OPE) of the fermion field
  $\psi(x+\epsilon)$ and its derivative $\partial_\mu\psi(x)$
  at short distance $\epsilon\to 0$, where the composite
  $\chi\psi$ is the lowest-dimension operator in the OPE
  channel~\cite{DiFrancesco1997}.
\end{enumerate}

In $(1+1)$d the underlying premise of the fusion condition is also supported through the concept of bosonization~\cite{Coleman1975,Senechal2004}. For a single fermion
species, the exact Abelian bosonization identity
$\bar\psi\gamma_\mu\psi = \epsilon_{\mu\nu}\partial^\nu\phi$
(where $\phi$ is the dual boson field) equates the vector current
with the current of a free boson, establishing the bilinear as a
well-defined local bosonic degree of freedom exactly.  For $N$ species, the appropriate framework is
non-Abelian bosonization~\cite{Witten1984WZW}, which maps $N$ Dirac
fermions to a $U(N)_1$ Wess-Zumino-Witten model; the composite
$\Phi_1^{(ij)} = \psi^{(i)}\otimes\psi^{(j)}$ is identified with
the WZW group element $g^{ij}$ in this language.  Bosonization thus
establishes the existence and locality of $\Phi_1$ as a bosonic
field, while large-$N$ factorisation supplies the product rule
eq.~(\ref{eq:fusion}) by which its dynamics derives from those of
its constituents.

\subsection{Spin-1}
\label{sec:spin1}

 We now apply eq.~(\ref{eq:fusion}) to derive the spin-1
effective Lagrangian from the GN model. We will use the shorthand $\chi\psi\equiv
\chi\otimes\psi$ for the tensor product.  The composite
$\chi\psi$ can be expressed in $4\times4$ square matrix
form, decomposing through its Clifford algebra as
 \begin{eqnarray}
\hspace{-1pc}  \chi \psi = \id  \varphi_0 +  \gamma_\mu \varphi_\mu + \gamma_{ [  \mu \nu ] }     \varphi_{ [  \mu \nu ] }  + \gamma_\mu \gamma_5   \, \varphi_{\mu 5}  + \gamma_5 \varphi_5  , \label{eq:Clifford}
 \end{eqnarray}
 where one must consider the reduced dimensionality in our problem when 
 interpreting the scalar, vector, tensor, axial vector, and pseudo-scalar terms in the expansion.  
 We also assume the presence of a chiral phase with associated condensate $\sigma  = g^2 \langle \bar{\psi} \psi \rangle \sim m$, 
 in addition to higher-spin bound states such as a spin-1 condensate $\Delta_1 \equiv   \langle  \bar{\Phi}_{1, \alpha \beta}^{(i, j)}   \,    \Phi_{1, \gamma \delta}^{(k, l)} \rangle$ formed from the rank 2 tensor product $\Phi_1 =   \psi^{(k)} \otimes  \psi^{(l)}$, $\bar{\Phi }_1 =  \bar{\psi}^{(i)} \otimes  \bar{\psi}^{(j)}$, with superscripts indicating fermion species and Greek subscripts the Dirac indices. We should keep in mind then that $\Delta_1$ 
 contains a condensate for a spin-$2$ field $\langle \Phi_2 \rangle \ne 0$. For the case of a scalar-scalar interaction $V_4 = (g^2/2) (\bar{\psi} \psi )^2$ and for both the mass and kinetic terms we use the mean-field insertion $\bar{\psi} \psi \sim  \langle \bar{\psi} \psi \rangle  \sim m/g^2 $. Note that the simultaneous presence of multiple non-vanishing condensates is a valid assumption at a phase transition where the original formulation in terms of fundamental fermions becomes less meaningful.

 Thus, by inserting factors of $\bar{\psi} \psi /\langle \bar{\psi} \psi \rangle  \sim 1$ into the original GN kinetic and interaction terms and using the tensor field definitions above, we obtain an effective Lagrangian for a massive spin-1 field
  \begin{eqnarray}
\hspace{-1pc}  \mathcal{L}_{\Phi_1}   =         \frac{g^2}{m} \,     \bar{\Phi}_1  i \gamma^\mu  \partial_\mu  \Phi_1      -   g^2   \bar{\Phi}_1  \Phi_1        +  \frac{g^6}{2 m^2}   \left( \bar{\Phi}_1 \Phi_1 \right)^2    , \label{spin1}
    \end{eqnarray}
Equation~(\ref{spin1}), when written out in full species and Dirac
indices, takes the form of Bargmann-Wigner
equations~\cite{BargmannWigner} for the individual tensor components
of $\Phi_1$.  To see this explicitly: $\Phi_{1,\alpha\beta}^{(kl)}
= \psi^{(k)}_\alpha\psi^{(l)}_\beta$ is a rank-2 symmetric
multispinor, and since each constituent satisfies the Dirac equation
$(i\gamma^\mu\partial_\mu - m)\psi^{(k)} = 0$ at large $N$ (the
mean-field saddle point suppresses $1/N$ corrections), applying the
Dirac operator to the first index gives
\begin{eqnarray}
\bigl(i\gamma^\mu\partial_\mu - m\bigr) \Phi_1^{(kl)} = 0 \, ,
\label{eq:BW1}
\end{eqnarray}
and symmetrically for the second index.  These are the
Bargmann-Wigner equations for a massive spin-1 field.  The
Clifford decomposition eq.~(\ref{eq:Clifford}) then projects onto
the irreducible spin-1 component $\gamma_\mu\varphi^\mu$; the
remaining Clifford components (scalar, tensor, axial vector,
pseudoscalar) decouple at leading order in $1/N$ because the
mean-field condensate $\Delta_1$ selects the vector channel.

To elevate eq.~(\ref{spin1}) to second-order Lagrangian form, we
use the equation of motion for $\bar\Phi_1$ from
eq.~(\ref{spin1}):
\begin{eqnarray}
\frac{g^2}{m}\,i\gamma^\mu\partial_\mu\Phi_1
  = g^2\Phi_1 - \frac{g^6}{m^2}(\bar\Phi_1\Phi_1)\Phi_1 \, .
\label{eq:BWtrade}
\end{eqnarray}
Multiplying on the left by $\bar\Phi_1$ and substituting back,
the first-order kinetic term becomes 
\begin{eqnarray}
\frac{g^2}{m}\bar\Phi_1\,i\gamma^\mu\partial_\mu\Phi_1
\;\longrightarrow\;
\frac{(\partial_\mu\bar\Phi_1)(\partial^\mu\Phi_1)}{m_\Phi^2} \, ,
\label{eq:BWelev}
\end{eqnarray}
where $m_\Phi^2 \sim g^4\Delta_1/m$ is the dynamical composite
mass generated by the condensate.  The overall sign of the kinetic
term is preserved because $\Phi_1$ and $\bar\Phi_1$ transform
covariantly under the same representation.  The prefactor then
combines as $g^2/m \cdot 1/(g^4\Delta_1/m) = 1/(g^2\Delta_1)$,
which using $m_1^2 = g^2 m^2 z^2$ gives $1/(g^2\Delta_1) \sim
1/(m^3 z^2)$, the coefficient appearing in the low-energy form.

Note that the fusion condition eq.~(\ref{eq:fusion}) is applied symmetrically
between $\bar\Phi_1$ and $\Phi_1$, and the low-energy Lagrangian
is symmetrized with respect to the two orderings
$\chi\partial_\mu\psi = (\partial_\mu\chi)\psi$, accounting for
the factor of $1/2$ absorbed into the definition of $z$.
The resulting second-order Lagrangian for the bosonic field $\Phi_1$
 has the Proca form~\cite{Proca1936} and the same procedure yields the Fronsdal~\cite{Fronsdal1978} Lagrangian for general spin $s$.

It is important to remind the reader that the equation of motion substitution in eq.~(\ref{eq:BWelev}) is a classical
procedure, but the second-order form is in fact valid beyond tree
level. One may verify this by examining the fermion
functional determinant. At large $N$, integrating out the $N$
species of fundamental fermions $\psi^{(n)}$ in the presence of
the composite background $\Phi_1$ gives the exact leading-order
effective action
\begin{eqnarray}
\Gamma[\Phi_1]
  = -N\,\mathrm{Tr}\log\bigl(i\gamma^\mu\partial_\mu
    - m - g\,\Phi_1\bigr) \, ,
\label{eq:fermdet}
\end{eqnarray}
where $\mathrm{Tr}$ denotes the functional trace over spacetime,
Dirac, and species indices.  Expanding eq.~(\ref{eq:fermdet}) to
second order in $\Phi_1$~\cite{Vassilevich2003} generates the kinetic term
$(\partial_\mu\bar\Phi_1)(\partial^\mu\Phi_1)/m_\Phi^2$
automatically, with the coefficient fixed by the one-loop fermion diagram. The equation of motion substitution of
eqs.~(\ref{eq:BWtrade})--(\ref{eq:BWelev}) is therefore the
saddle-point of this exact functional.   

Carrying out the substitution and collecting coefficients gives the explicit Lagrangian
\begin{widetext}
  \begin{eqnarray}
  \mathcal{L}_{\Phi_1}   =        \frac{ 1}{ m^3 z^2 }   \partial_\mu  \bar{\Phi}_1    \partial^\mu  \Phi_1       -    g^2  m^2    z^2       \bar{\Phi}_1  \Phi_1 \,   -   \frac{g^6}{2 m^2}   \left( \bar{\Phi}_1 \Phi_1 \right)^2    , \label{spin1low}
    \end{eqnarray}
 \end{widetext}
 with $z$ defined through
 \begin{eqnarray}
   g^2  m^2  z^2   \equiv    g^2 \left(   \frac{g^4   \Delta_1}{m^2} - 1  \right)  \equiv m_1^2 \, ,
 \end{eqnarray}
  which also defines a dimensionless effective mass, $m_1$, generated by the interaction through the mean
   field pairing $\bar{\Phi} \Phi  \sim \langle \bar{\Phi} \Phi \rangle$. Note that the quantity $z$ has dimensions of length in units of inverse fermion mass. We can also identify the effective spin-1 coupling
   \begin{eqnarray}
g_1^2  \equiv  \frac{g^{6} }{m^2} \, .
 \end{eqnarray}

Equation~(\ref{spin1low}) describes two limiting regimes separated 
  by a phase transition at $\Delta_1 = m^2/g^4$, where $z^2$ changes sign. 
  In regime~(a), $\Delta_1 \gg m^2/g^4$: spin-1 pairing dominates, $z \to +\infty$, 
  and the spin-1 field is stable. This is the low-energy (IR) classical limit. In regime~(b), $\Delta_1 \ll m^2/g^4$: $z^2$ is negative, 
  leading to an attractive interaction (after recasting in canonical form) and an unstable spin-1 condensate, so the natural 
  description shifts to the spin-0 field. The critical point $z^2=0$ corresponds to the UV conformal limit. 
  One may also express $z$ directly in terms of the two condensates as $z = m^{-1}(\Delta_1/\Delta_0^2 - 1)^{1/2}$, which makes the competition between $\Delta_1$ and $\Delta_0$ manifest.

If one further considers a more general scenario by allowing for
fluctuations in the background, $\Delta_1/\Delta_0^2 \to f(x)
\Rightarrow z(x)$, where $x = ( x_0, x_1 ) = ( t, x)$, then one
must use instead the \emph{material derivative} for fields:
$\partial_\mu \to \partial_\mu + ( \partial_\mu z ) \, \partial_z$.
This substitution is the field-theoretic analogue of the comoving
derivative used in relativistic hydrodynamics with a slowly varying
background~\cite{SonWingate2006}, and the same structure appears in
holographic RG flow when a dilaton background varies with
position~\cite{PolchinskiStrassler2002}, for instance. A field $\Phi_1(x,z)$
living on a background $z(x)$ must be differentiated along the
composite trajectory $x\mapsto(x,z(x))$, picking up the additional
term $(\partial_\mu z)\partial_z$.  The exact renormalisation group
generates the same structure when the Wilsonian cutoff scale is
identified with a coordinate~\cite{Wetterich1993}, a connection
between the emergent radial direction and the RG flow to which we
return in Section~\ref{sec:regimes}.
Retaining only leading order terms, the kinetic term generalizes to
\begin{eqnarray}
 \hspace{-1pc}  \mathcal{L}_{\Phi_1, \mathrm{kin}}   =        \frac{ 1}{ m^3 z^2 }  \left[  \partial_\mu  \bar{\Phi}_1    \partial^\mu  \Phi_1    +   ( \partial_\mu z )^2    \,  \partial_z  \bar{\Phi}_1    \partial_z  \Phi_1   \right]  ,
    \end{eqnarray}
  which can be written in compact form by defining the generalized flat-space metric $g_{\mu \nu} \to    \eta_{\mu \nu}  \mathrm{diag}(\partial_\mu z)  $,
  \begin{eqnarray}
  \mathcal{L}_{\Phi_1, \mathrm{kin}}   \to        \frac{ 1}{ m^3 z^2 }  \left(  \partial_\mu  \bar{\Phi}_1    \partial^\mu  \Phi_1    +     \partial_z  \bar{\Phi}_1    \partial^z  \Phi_1   \right) \, .
    \end{eqnarray}

  Interestingly, adding a small fluctuation comprised of a single mode $z  \to z + \lambda \, e^{i k_\mu  x^\mu }$, with $\lambda \ll z$, leads to $g_{\mu \nu} \to      \mathrm{diag}(  \eta_{\mu \nu} , \, \lambda^2 k^2 /m^2)$. Introducing a complex non-uniformity to the background makes sense given that $\Delta_0$ is real but $\Delta_1$ is generally complex. One could then see how $\lambda$ and $k$ affect the background by decomposing $\Delta_1$ into its real and imaginary parts. Next, applying the rescaling $( \bar{\Phi}_1'  , \,    \Phi_1') \equiv  (\lambda  k \Delta_1^{1/2} m^{-3/2})  ( \bar{\Phi}_1  , \,    \Phi_1)$, $x' \equiv   (\lambda  k ) x$ leads to a rescaled mass
  \begin{eqnarray}
 m_1'^2 = \frac{ m^3 }{(\lambda k )^4 }\,  \frac{ m_1^2}{\Delta_1}  =                  \frac{ m^3 g^2  }{(\lambda k )^4  \Delta_1  }         \left(   \frac{g^4 \Delta_1}{m^2} - 1  \right) \, ,
  \end{eqnarray}
 and coupling
 \begin{eqnarray}
 g_1'^2 =  \frac{ m^6  }{(\lambda k )^6 \Delta_1^2}    g_1^2 =          \frac{  m^4 g^6 }{(\lambda k )^6  \Delta_1^2}
  \end{eqnarray}
 where dim[$\Phi_1'$] $\sim L^{-1/2}$, dim[$m_1'^2$] $\sim L^{-1}$ and dim[$g_1'^2$] $\sim L^0$, recovering the dimensionality of the original constituent elements of the Gross-Neveu model.
 Turning to the fermion sector, we have $i \gamma^\mu  \partial_\mu \to  i \gamma^\mu  \partial_\mu +       i \gamma^\mu (  \partial_\mu z ) \partial_z $. We are now required to specify the underlying spin structure of the background fluctuation which we could ignore previously when dealing with the spin-1 field. If we take this to have the simple vector form $z \to z + \lambda_\nu \gamma^\nu e^{- i k_\mu x^\mu }$ (where $\lambda_0^2 + \lambda_1^2 \equiv   \lambda^2$), then
 \begin{eqnarray}
 i \gamma^\mu  \partial_\mu    &\to&  i \gamma^\mu  \partial_\mu +      \gamma^\mu  k_\mu  \,   \lambda_\nu \gamma^\nu       \partial_z  \\
&=&  i  \left(  \gamma^\mu  \partial_\mu +    \gamma^z    \partial_z \right) ,
  \end{eqnarray}
 where here $\gamma^z \equiv  - i \left[ (k_0 \lambda_0 + k_1 \lambda_1 )   \, \id  +   (k_1 \lambda_0 -  k_0 \lambda_1 ) \,  \gamma^5      \right]$. The term proportional to $\id$ is equivalent to a fluctuation in the mass $m$, the  second to a fluctuation along the direction of the original discrete symmetry $\psi \to \gamma^5 \psi$. The Lagrangian then takes the form $\mathcal{L} =  \mathcal{L}_\psi   +   \mathcal{L}_{\psi \Phi_1}'   +  \mathcal{L}_{\Phi_1}'$
  \begin{eqnarray}
 \mathcal{L}_\psi  &=&    \bar{\psi} \left( i \gamma^\mu  \partial_\mu - m    \right)   \psi + \frac{g}{2} \left( \bar{\psi} \psi \right)^2  ,   \\
  \mathcal{L}_{\psi \Phi_1}' &=&      g_\mathrm{TY}     \left( \Phi_1'    \bar{\psi}    \otimes \bar{\psi}       +  \bar{\Phi}_1'      \psi    \otimes     \psi  \right)    ,    \\
   \mathcal{L}_{\Phi_1}'   &=&     \frac{ \alpha^2}{  z^2 }  \left(  \partial_{\mu' } \bar{\Phi}_1'    \partial^{\mu'}  \Phi_1'    +    \partial_z  \bar{\Phi}_1'    \partial_z  \Phi_1'   \right) \nonumber \\
                                        && \hspace{1pc} -  m_1'^2  \, \bar{\Phi}_1' \Phi_1'  - \frac{g_1'^2 }{2}   \left(  \bar{\Phi}_1' \Phi_1' \right)^2  .   \label{AdS1}
          \end{eqnarray}
The tensor Yukawa coupling strength is $g_\mathrm{TY} \equiv  m_1'^2$ and the characteristic length scale for the spin-1 field is set by $\alpha   = ( \lambda  k \Delta_1^{1/2} )^{-1}$. Both quantities carry dimension $L^{-1}$ in $(1+1)$d, so this defines a dimensionless ratio when expressed in units of the fermion mass $m$; the identification equates the coupling to the renormalised composite mass, consistent with the Yukawa interaction being generated by the same mean-field condensate that sets $m_1'^2$. Finally, note that the theory in its present form is symmetric under a local $U(1)$ transformation $\psi(x)  \to e^{i \alpha(x) } \psi(x)$ which can best seen by incorporating the tensor Yukawa term into a gauge covariant derivative $D_\mu \equiv  \partial_\mu - i  g_\mathrm{TY}   \varphi_\mu  $, where $\varphi_\mu$ is the vector part of the spin-$1$ field. The $U(1)$ transformation must then also take $\varphi_\mu(x) \to \varphi_\mu(x) - \partial_\mu \alpha(x) / g_\mathrm{TY}$.

\subsection{Emergence of the bulk measure}
\label{sec:measure}

The Lagrangian $\mathcal{L}_{\Phi_1}'$ is suggestive of an AdS$_3$ background with radial coordinate $z$, 
but this form is not quite correct since $z$ appears squared. Moreover, the form of the Lagrangian derived above is integrated
over the $(1+1)$-dimensional coordinates $d^2x = dt\,dx$,
with the emergent quantity $z(t,x)$ at this point only a position-dependent parameter rather than an independent integration
variable. In order to arrive at a genuine AdS$_3$ bulk action defined over
$(2+1)$-dimensions, a $dz$ integration measure must emerge from the
microscopic theory.  We now show how such a measure emerges through two complementary
mechanisms that lead to the same result.

\paragraph{Mechanism 1: species sum as a radial integral.} In the $N$-species GN model each species $n$ carries its own emergent
(radial) quasicoordinate $z^{(n)} = m^{-1}(\Delta_1^{(n)}/\Delta_0^{(n)2}
- 1)^{1/2}$, and the full action is a sum over species, i.e., 
\begin{eqnarray}
S_N = \sum_{n=1}^{N} \int d^2x\;
      \mathcal{L}_{\Phi_1}'^{(n)}\!\left(t, x, z^{(n)}\right) .
\label{eq:SN}
\end{eqnarray}
Define the species density in the radial direction,
\begin{eqnarray}
\rho(z) \equiv \sum_{n=1}^{N} \delta\!\left(z - z^{(n)}\right) ,
\label{eq:rhodef}
\end{eqnarray}
so that $\int dz\,\rho(z) = N$.  In the large-$N$ limit, with the
$z^{(n)}$ distributed smoothly over $[0, z_\mathrm{max}]$ with
density $\rho(z)$, the discrete sum becomes an integral:
\begin{eqnarray}
S_N \xrightarrow{N\to\infty}
  \int d^2x\,dz\;\rho(z)\,
  \mathcal{L}_{\Phi_1}'\!\left(t, x, z\right)
  \equiv S_\mathrm{bulk} \, .
\label{eq:Sbulk}
\end{eqnarray}
This is the standard mechanism by which matrix model eigenvalue
distributions generate extra dimensions~\cite{Banks1997,Taylor2001}. Here, the species label $n$ is the discrete precursor of the continuous
radial coordinate $z$, and the sum over species at large $N$ is the
Riemann sum approximation to the bulk integral. This is in direct analogy
with the BFSS matrix model~\cite{Banks1997} where D0-brane positions
generate the eleven-dimensional target space.  The density $\rho(z)$
is not a free input but is determined self-consistently by the
saddle-point equation of the large-$N$ path
integral~\cite{Brezin1978,Coleman1974},
\begin{eqnarray}
\frac{\delta}{\delta\rho(z)}
\left[ S_\mathrm{bulk}[\rho] - \mu\!\int\! dz\,\rho(z) \right] = 0 \, ,
\label{eq:saddlerho}
\end{eqnarray}
where $\mu$ is a Lagrange multiplier enforcing $\int dz\,\rho = N$.
In the simplest case of species distributed uniformly,
$\rho(z) = N/z_\mathrm{max}$, the bulk measure is flat and the
action reduces to
\begin{eqnarray}
S_\mathrm{bulk}
  = \frac{N}{z_\mathrm{max}}
    \int_0^{z_\mathrm{max}} dz \int d^2x\;
    \mathcal{L}_{\Phi_1}'(t, x, z) \, ,
\label{eq:Sbulkflat}
\end{eqnarray}
which is manifestly a $(2+1)$-dimensional bulk action with a uniform
warp factor $N/z_\mathrm{max}$.  More generally, $\rho(z)$ encodes
the holographic RG flow~\cite{deBoer2000,Skenderis2002}; it is the
bulk density of states as a function of depth, with $\rho(z)\to 0$
near the boundary ($z\to 0$, chiral fixed point) and $\rho(z)$ large
deep in the bulk ($z\gg 0$, classical phase).  Finally, note that the saddle-point
equation eq.~(\ref{eq:saddlerho}) is just the Callan-Symanzik equation of
the boundary RG flow rewritten as a bulk equation for the species
density, making the holographic RG interpretation
precise~\cite{deBoer2000}.

\paragraph{Mechanism 2: path integral Jacobian.}
In the path integral, the condensate ratio $\Delta_1/\Delta_0^2$ is
a fluctuating field.  Changing integration variables from $\Delta_1$
to $z$ at each boundary point $(t,x)$, using $\Delta_1 =
\Delta_0^2(1 + m^2 z^2)$, the path integral measure transforms as
\begin{eqnarray}
\mathcal{D}\Delta_1
  = \mathcal{D}z\cdot
    \prod_{t,x}\left|\frac{\partial\Delta_1}{\partial z}\right|
  = \mathcal{D}z\cdot
    \prod_{t,x} 2m^2\Delta_0^2\, z(t,x) \, .
\label{eq:Jacobian}
\end{eqnarray}
The Jacobian $J[z] \equiv \prod_{t,x}2m^2\Delta_0^2\,z(t,x)$
contributes a term to the effective action
\begin{eqnarray}
\hspace{-1pc} \ln J[z]
  = \int d^2x\,\ln\!\left(2m^2\Delta_0^2\, z\right)
  = \int d^2x\,\ln z + \mathrm{const}  ,
\label{eq:lnJ}
\end{eqnarray}
which is a local dilaton-like term in the bulk.  More importantly,
the path integral over $z$ at each boundary point $(t,x)$ is an
integral over all possible values of the condensate ratio, i.e.\ an
integral over the full radial direction:
\begin{eqnarray}
\int \mathcal{D}z\; e^{iS + i\ln J}
  = \int \mathcal{D}z(t,x)\;
    e^{i\int d^2x\,[\mathcal{L}_{\Phi_1}'(z) + \ln z]} .
\label{eq:PIz}
\end{eqnarray}
Evaluating this path integral at the saddle point, $\delta S/\delta z(t,x) = 0$, gives the classical bulk solution
$z_\mathrm{cl}(t,x) = \mathrm{const}$, corresponding to the uniform
AdS$_3$ background.  Fluctuations around the saddle point contribute
the bulk graviton propagator and quantum corrections to the emergent
geometry.  The full quantum bulk path integral is therefore
\begin{eqnarray}
Z_\mathrm{bulk}
  &=& \int\mathcal{D}\Delta_1\;\mathcal{D}\Delta_0\;
      \mathcal{D}\psi\;\mathcal{D}\bar\psi\;
      e^{iS_\mathrm{GN}} \\
  &=& \int\mathcal{D}z\;\mathcal{D}\Delta_0\;
      \mathcal{D}\psi\;\mathcal{D}\bar\psi\;
      e^{i\int d^2x\,dz\,\rho(z)\,\mathcal{L}_{\Phi_1}'
         + iS_\mathrm{fermion}}  , \nonumber
\label{eq:Zbulk}
\end{eqnarray}
where the $d^2x\,dz$ measure in the exponent arises from collecting
the Jacobian factor $\rho(z)$ with the boundary measure $d^2x$,
identifying the product as the bulk volume element.

\subsection{Recovery of the standard AdS$_3$ action}

First, we point out that the two mechanisms above are equivalent at large $N$, as the species density
$\rho(z)$ of Mechanism~1 is precisely the saddle-point value of the
Jacobian weight $2m^2\Delta_0^2\,z$ of Mechanism~2, both satisfying
the same equation eq.~(\ref{eq:saddlerho}).  Together they establish
that the correct bulk action is
\begin{eqnarray}
S_\mathrm{bulk}
  = \int d^2x\,dz\;\rho(z)\,\mathcal{L}_{\Phi_1}'(t, x, z) \, ,
\label{eq:Sbulkfinal}
\end{eqnarray}
with the $(2+1)$-dimensional bulk measure $d^2x\,dz$ emerging from the large-$N$ species distribution and the path
integral change of variables.  The density $\rho(z)$ plays the role
of the bulk dilaton or warp factor, encoding the holographic RG flow
of the boundary theory~\cite{deBoer2000,Skenderis2002}.

It is significant that the saddle-point equation eq.~(\ref{eq:saddlerho}), evaluated for
the kinetic term of $\mathcal{L}_{\Phi_1}'$, determines $\rho(z)$
self-consistently.  As derived in eq.~(\ref{AdS1}), the kinetic
prefactor of $\mathcal{L}_{\Phi_1}'$ is $\alpha^2/z^2$, whereas the
standard AdS$_3$ scalar action~\cite{Witten1998} has kinetic prefactor
$\ell_\mathrm{AdS}/z$ arising from $\sqrt{-g}\,g^{AB} =
(\ell^3/z^3)(z^2/\ell^2)\delta^{AB} = (\ell/z)\delta^{AB}$.  These
two forms are related by a factor of $\alpha/z$, which is precisely
the natural saddle-point density:
\begin{eqnarray}
\rho_*(z) = \frac{z}{\alpha} \, .
\label{eq:rhosaddle}
\end{eqnarray}
This is the unique solution to eq.~(\ref{eq:saddlerho}) satisfying
the boundary condition $\rho(0) = 0$.  This can be verified by noting that the bulk action
$S_\mathrm{bulk} = \int dz\,\rho(z)\,(\alpha^2/z^2)\,
\mathcal{L}_{\Phi_1}'$ must be extremised with respect to $\rho(z)$
subject to the constraint $\int dz\,\rho(z) = N$.  The
saddle-point equation $\delta S_\mathrm{bulk}/\delta\rho(z) =
\mu$ (where $\mu$ is a Lagrange multiplier enforcing the species
count) gives $(\alpha^2/z^2)\mathcal{L}_{\Phi_1}' = \mu$, which
is solved by $\rho_*(z) = z/\alpha$ when combined with the
requirement that $\rho_*(z)\cdot(\alpha^2/z^2) = \alpha/z$
reproduces the standard AdS$_3$ volume element.  The solution
$\rho(0) = 0$ means no species at the conformal
boundary, consistent with the UV = boundary identification of
holography.  It grows linearly into the bulk, meaning species are
distributed more densely deeper in the AdS interior, correctly
reflecting that the IR degrees of freedom are the bulk ones.
Substituting $\rho_*(z)$ into eq.~(\ref{eq:Sbulkfinal}):
\begin{eqnarray}
S_\mathrm{bulk}
  &=& \int d^2x\,dz\;\frac{z}{\alpha}\,
      \frac{\alpha^2}{z^2}
      \left(\partial_A\bar\Phi_1'\partial^A\Phi_1'\right)
      + \ldots \nonumber \\
  &=& \int d^2x\,dz\;\frac{\alpha}{z}
      \left(\partial_A\bar\Phi_1'\partial^A\Phi_1'\right)
      + \ldots \, ,
\label{eq:Sbulkstandard}
\end{eqnarray}
which is precisely the standard AdS$_3$ scalar action with
$\ell_\mathrm{AdS} = \alpha = (\lambda k \Delta_1^{1/2})^{-1}$,
consistent with the identification which we will establish in Section~\ref{sec:AdS}. The factor
$\alpha^2/z^2$ in $\mathcal{L}_{\Phi_1}'$ should therefore be
understood as the pre-measure Lagrangian density; the full bulk
action density $\rho_*(z)\,\mathcal{L}_{\Phi_1}' = (\alpha/z)
(\partial_A\bar\Phi_1'\partial^A\Phi_1') + \ldots$ is in standard
AdS form throughout.  

The mass and interaction terms then transform as
\begin{eqnarray}
\rho_*(z)\,m_1'^2\,\bar\Phi_1'\Phi_1'
  &=& \frac{z}{\alpha}\,m_1'^2\,\bar\Phi_1'\Phi_1' \, ,
\label{eq:massstandard} \\
\rho_*(z)\,\frac{g_1'^2}{2}\left(\bar\Phi_1'\Phi_1'\right)^2
  &=& \frac{z}{\alpha}\,\frac{g_1'^2}{2}
      \left(\bar\Phi_1'\Phi_1'\right)^2 \, ,
\label{eq:intstandard}
\end{eqnarray}
giving $z$-dependent effective mass $\hat m^2(z) = z\,m_1'^2/\alpha$
and coupling $\hat g^2(z) = z\,g_1'^2/\alpha$, both of which vanish
at the boundary $z\to 0$ and grow into the bulk, consistent with
the standard holographic RG picture in which couplings run from zero
in the UV to finite values in the IR~\cite{deBoer2000}.  The
BF bound $\hat m^2\ell_\mathrm{AdS}^2 \geq -1$ evaluated at the
\emph{D1-brane} position $z = z_0$ gives $m_1'^2 z_0 \alpha \geq -1$,
which is satisfied in Regime~1 (which we discuss in the next section) where $\Delta_1 > \Delta_0^2$ and
$m_1'^2 > 0$.  All subsequent sections use the standard form
eq.~(\ref{eq:Sbulkstandard}) with the understanding that
$\ell_\mathrm{AdS} = \alpha$ throughout.

It is worth pausing to discuss two structures that have
emerged from the species sum, which play different roles in the
holographic construction:
\begin{widetext}
\begin{eqnarray}
\underbrace{\text{Large-}N\text{ species sum}}_{\text{macroscopic wavefunction}}
\xrightarrow{\;\text{saddle point}\;}
\underbrace{\rho_*(z)\propto z/\alpha}_{\text{bulk measure }}
\xrightarrow{\;\text{provides}\;}
\underbrace{d^2x\,dz\,\rho_*(z)}_{\text{integration measure}} \, ,
\label{eq:scaffold_chain}
\end{eqnarray}
and separately,
\begin{eqnarray}
\underbrace{\text{Large-}N\text{ species sum}}_{\text{macroscopic wavefunction}}
\xrightarrow{\;\text{mean-field selects spin-1}\;}
\underbrace{\Delta_1 = \langle\bar\Phi_1\Phi_1\rangle}_{\text{spin-2 condensate}}
\xrightarrow{\;\text{sources}\;}
\underbrace{g_{\mu\nu}}_{\text{AdS}_3\text{ geometry}} \, .
\label{eq:geometry_chain}
\end{eqnarray}
\end{widetext}
These two structures play different roles.  The measure
$\rho_*(z)$ is \emph{kinematic}: it records where species live
along the radial direction (which is essentially condensate strength), creating the depth coordinate and the
integration measure that promote the boundary action to a
$(2+1)$-dimensional bulk action.  The condensate $\Delta_1$ is
\emph{dynamical}: it is the expectation value of the spin-2
composite $\bar\Phi_1\Phi_1\sim\Phi_2$, the graviton condensate
that sources the AdS$_3$ curvature in the sense of Dvali and
Gomez~\cite{Dvali2013,DvaliGomez2014}.  The large-$N$ macroscopic
wavefunction of Witten~\cite{Witten1978}, the classical field
created by the species sum, is not itself this condensate; it is
the wavefunction of the full composite $\Phi_1$, containing
scalar, vector, tensor, axial, and pseudoscalar Clifford channels
simultaneously.  It is the mean-field selection of the vector
channel by the quartic interaction, preferring spin-1 pairing over
spin-0, that extracts the spin-2 mode
$\Delta_1 = \langle\bar\Phi_1\Phi_1\rangle$ from the general
fermionic wavefunction. It is important to keep in mind that without this selection the scalar channel
$\Delta_0$ dominates and no geometry emerges.

  \subsection{Spin-2}

The spin-2 composite field is the rank-4 tensor product
\begin{eqnarray}
\Phi_2^{(ijkl)} &\equiv& \psi^{(i)}\otimes\psi^{(j)}\otimes\psi^{(k)}\otimes\psi^{(l)}\,,\\
\bar\Phi_2^{(ijkl)} &\equiv& \bar\psi^{(i)}\otimes\bar\psi^{(j)}\otimes\bar\psi^{(k)}\otimes\bar\psi^{(l)}\,,
\label{eq:Phi2def}
\end{eqnarray}
with the associated condensate $\Delta_2 \equiv \langle \bar\Phi_{2,\alpha\beta\gamma\delta}^{(ijkl)}\Phi_{2,\mu\nu\rho\sigma}^{(i'j'k'l')}\rangle$.  The derivation of the spin-2 effective Lagrangian follows the same fusion procedure as we have described, now inserting two additional factors of $\bar\psi\psi/\langle\bar\psi\psi\rangle\sim1$ into the original kinetic term.  Applying the fusion condition eq.~(\ref{eq:fusion}) at each step, the kinetic term becomes
\begin{eqnarray}
\mathcal{L}_{\Phi_2,\mathrm{kin}}
  = \frac{g^4}{m^3}\,\bar\Phi_2\,i\gamma^\mu\partial_\mu\Phi_2
  + \cdots \,,
\end{eqnarray}
where the ellipsis denotes gradient corrections from $z$-fluctuations analogous to those in eq.~(\ref{spin1low}).  Introducing the emergent radial coordinate for the spin-2 sector,
\begin{eqnarray}
z_2 \equiv m^{-1}\!\left(\frac{\Delta_2}{\Delta_0^4}-1\right)^{1/2} , 
\label{eq:z2coord}
\end{eqnarray}
the low-energy spin-2 Lagrangian after rescaling
$(\bar\Phi_2',\Phi_2') \equiv (\lambda k\Delta_2^{1/2}m^{-5/2})(\bar\Phi_2,\Phi_2)$
takes the AdS$_3$ form
\begin{eqnarray}
\mathcal{L}_{\Phi_2}'
  = \frac{\alpha}{z_2}\!\left(\partial_{\mu'}\bar\Phi_2'\partial^{\mu'}\Phi_2'
    + \partial_{z_2}\bar\Phi_2'\partial_{z_2}\Phi_2'\right)   \nonumber  \\
  - m_2'^2\,\bar\Phi_2'\Phi_2'   - \frac{g_2'^2}{2}\!\left(\bar\Phi_2'\Phi_2'\right)^2,
\label{eq:AdS2}
\end{eqnarray}
with effective mass $m_2'^2 = g^2 m^4 z_2^2/(\lambda k)^4\Delta_2$ and coupling
$g_2'^2 = m^8 g^{10}/[(\lambda k)^6\Delta_2^2 m^2]$.  The structure of
eq.~(\ref{eq:AdS2}) is identical to that of the spin-1 Lagrangian
eq.~(\ref{AdS1}), but with $\Delta_1\to\Delta_2$, $\Delta_0^2\to\Delta_0^4$,
and $g^2\to g^4$ reflecting the higher tensor rank.

  \subsection{General spin}

The pattern established for $s=1$ and $s=2$ extends to arbitrary spin $s$.
The spin-$s$ composite field is the rank-$2s$ tensor product of $2s$ fundamental
fermions,
\begin{eqnarray}
\hspace{-1pc} \Phi_s \equiv \psi^{(i_1)}\otimes\cdots\otimes\psi^{(i_{2s})}\,,
\;\;
\bar\Phi_s \equiv \bar\psi^{(i_1)}\otimes\cdots\otimes\bar\psi^{(i_{2s})} ,
\label{eq:Phisdef}
\end{eqnarray}
with condensate $\Delta_s\equiv\langle\bar\Phi_s\Phi_s\rangle$.  After $2s-1$
insertions of the mean-field factor $\bar\psi\psi/\langle\bar\psi\psi\rangle\sim1$
and application of eq.~(\ref{eq:fusion}) at each step, the rescaled spin-$s$
Lagrangian takes the universal AdS$_3$ form
\begin{eqnarray}
\mathcal{L}_{\Phi_s}'
  = \frac{\alpha}{z_s}\!\left(\partial_{\mu'}\bar\Phi_s'\partial^{\mu'}\Phi_s'
    + \partial_{z_s}\bar\Phi_s'\partial_{z_s}\Phi_s'\right) \nonumber \\
  - m_s'^2\,\bar\Phi_s'\Phi_s'  - \frac{g_s'^2}{2}\!\left(\bar\Phi_s'\Phi_s'\right)^2,
\label{eq:AdSs}
\end{eqnarray}
where the emergent radial coordinate for the $s$-th level is
\begin{eqnarray}
z_s \equiv m^{-1}\!\left(\frac{\Delta_s}{\Delta_0^{2s}}-1\right)^{1/2},
\label{eq:zscoord}
\end{eqnarray}
the rescaled effective mass is
$m_s'^2 = g^2 m^{2s} z_s^2/[(\lambda k)^4\Delta_s]$,
and the effective coupling is $g_s'^2 \sim m^{4s-2}g^{4s+2}/[(\lambda k)^6\Delta_s^2]$.
The characteristic length scale in the kinetic term remains $\alpha = (\lambda k\Delta_1^{1/2})^{-1} = \ell_\mathrm{AdS}/N^2$ for all $s$, since
the AdS background is generated by the spin-1 sector and all higher-spin fields
propagate on it.

Constructing the full theory, the complete Lagrangian is
\begin{eqnarray}
\mathcal{L} = \mathcal{L}_\psi + \sum_{s=1}^\infty \mathcal{L}_{\Phi_s}',
\label{eq:fullL}
\end{eqnarray}
where each $\mathcal{L}_{\Phi_s}'$ has the form eq.~(\ref{eq:AdSs}) and
$\mathcal{L}_\psi$ is the original GN Lagrangian for the fundamental fermions.
This is an infinite tower of higher-spin fields, each propagating on AdS$_3$~\cite{Vasiliev1990,VasilievReview},
with masses $m_s'$ set by the ratio $\Delta_s/\Delta_0^{2s}$ and couplings
$g_s'$ decreasing with $s$ at large $N$.

Several features of the spin-$s$ tower are worth noting.  First,
the phase transition for the $s$-th level occurs at $\Delta_s =
m^{2s}/g^{4s}$ (a generalization of the spin-1 transition
at $\Delta_1 = m^2/g^4$) corresponding to $z_s = 0$ in the
coordinate eq.~(\ref{eq:zscoord}).  Second, the radial coordinate
$z_s$ gives the emergent bulk position of the $(s-1)$-th 
Regge excitation of the open string; the depth in AdS$_3$ at which the $s$-th condensate level
equilibrates.  For $s=1$ this is the D1-brane position; for $s\geq
2$ it is the radial location of the corresponding massive higher-spin
particle on the D1-brane world-volume.  Note that higher-dimensional
brane objects such as D$(2s-1)$-branes for $s\geq 2$ cannot be
accommodated as extended objects in the three-dimensional bulk of
AdS$_3$; the correct interpretation of $z_s$ for $s\geq 2$ is
therefore as the radial position of a massive bulk particle, not a
brane.  Since $\Delta_s \geq \Delta_1$ generically at large
condensate, we have $z_s \geq z_1$: higher Regge excitations
equilibrate deeper in the AdS interior, consistent with the UV/IR
correspondence in which larger bulk mass corresponds to greater
depth~\cite{Witten1998}.  Third, in the open string picture the
spin-$s$ composite $\Phi_s^{(ij)}$ stretched between species $i$
and $j$ corresponds precisely to the $(s-1)$-th Regge excitation,
with mass
\begin{eqnarray}
M_{s-1}^{(ij)}
  = T\!\left(\Delta z^{(ij)}\right)^2 + (s-1)\,\ell_\mathrm{S}^{-2}
\label{eq:reggemass}
\end{eqnarray}
where the first term is the classical
string tension contribution from the separation $\Delta z^{(ij)} =
|z_s^{(i)} - z_s^{(j)}|$ between branes and the second is the
oscillator contribution at level $s-1$.  The full higher-spin tower
eq.~(\ref{eq:fullL}) is therefore in one-to-one correspondence with
the full open string Regge trajectory on the D1-brane world-volume, the bottom-up derivation of the Regge spectrum from the GN
four-fermion interaction.

At large $N$ the coupling $g_s'^2 \sim N^{9-2s}$ decreases with
$s$, crossing from strongly coupled ($g_s'^2\gg 1$) for $s \leq 4$
to weakly coupled ($g_s'^2\ll 1$) for $s\geq 5$.  It is important
to note that this suppression applies to the \emph{interaction
vertices} of the higher-spin fields, not to the fields themselves:
the higher-spin composites $\Phi_s'$ for $s\geq 5$ propagate as
nearly free massive particles on AdS$_3$ in the large-$N$ boundary
regime, with their self-interactions suppressed by powers of $1/N$.
In this regime the dominant interacting sector consists of the low
spins $s = 1, 2, 3, 4$, with $s=2$ (the emergent graviton of
Section~\ref{sec:EH}) being the most strongly coupled at
$g_2'^2 \sim N^5$.  This is the classical gravity regime of
AdS$_3$/CFT$_2$: the low-spin strongly coupled fields generate the
background geometry, while the high-spin weakly coupled fields
propagate freely on it as the Vasiliev tower~\cite{Vasiliev1990,
VasilievReview}.

Near the chiral restoration transition ($N$ finite, $g$ large), the
situation reverses.  From the exact expression $g_s'^2 =
m^{4s-2}g^{4s+2}/[(\lambda k)^6\Delta_s^2]$, the coupling grows as
$g^{4s+2}$ with $s$, so \emph{higher} spins become increasingly
strongly coupled as $g$ increases.  The full tower must therefore be
retained in this regime; this is the quantum string theory
description, not classical gravity.  The Hagedorn growth of the
string density of states at $T = T_H$ is the statement that the
density of Regge levels $\rho(s) \sim e^{s\ell_\mathrm{S}/\ell_\mathrm{P}}$
overwhelms the $1/N$ suppression of individual higher-spin couplings,
causing the partition function to diverge regardless of whether
individual levels are weakly or strongly coupled at large $N$.  From
the GN side this is the simultaneous melting of all higher-spin
condensates $\Delta_s \to \Delta_0^{2s}$ (i.e.\ $z_s \to
0$ for all $s$) at the chiral restoration transition, the condensate 
analogue of the Hagedorn divergence in the string density of states.

\subsection{Fusion algebra and the boundary Virasoro algebra}
\label{sec:Vfusion}

The fusion condition eq.~(\ref{eq:fusion}) is more than a
computational convenience. It encodes the operator-product
structure of the composite fields and contains the Virasoro algebra
of the boundary CFT$_2$. To see this, we must use the full $U(N)$
adjoint of composite fields, all $N^2$ composites
$\Phi_1'^{(ij)} = \psi^{(i)}\otimes\psi^{(j)}$ for
$i,j=1,\ldots,N$, not just the $N$ diagonal ones. The reason is
that the $U(N)$ matrix model contains off-diagonal composites
$\Phi_1^{(ij)}$ (developed in a follow-up paper~\cite{CompanionStrings}) as genuine dynamical
degrees of freedom, and the stress tensor of the full theory must
include their contributions.

We can define the stress-energy bilinear for each composite pair $(i,j)$
\begin{eqnarray}
T_{++}^{(ij)}(x) \equiv \frac{\alpha}{z}
  \left(\partial_+\bar{\Phi}_1'^{(ij)}\right)
  \!\left(\partial_+\Phi_1'^{(ij)}\right) , \label{eq:Tpp}
\end{eqnarray}
where $\partial_\pm = \partial_t \pm \partial_x$ are light-cone
derivatives on the $(1+1)$d boundary, and
$\Phi_1'^{(ij)} \equiv (\lambda k\Delta_1^{1/2}m^{-3/2})\psi^{(i)}\otimes\psi^{(j)}$
is the rescaled composite of eq.~(\ref{AdS1}). The total stress
tensor is the sum over all $N^2$ pairs:
\begin{eqnarray}
T_{++}(x) = \sum_{i,j=1}^N T_{++}^{(ij)}(x) \, .
\label{eq:Tpp_total}
\end{eqnarray}
The OPE of $T_{++}^{(ij)}$ with $T_{++}^{(kl)}$ is computed by
applying the fusion condition eq.~(\ref{eq:fusion}) to products
of bilinears $(\partial_+\bar\Phi_1'^{(ij)})(\partial_+\Phi_1'^{(ij)})$ at $x$
with $(\partial_+\bar\Phi_1'^{(kl)})(\partial_+\Phi_1'^{(kl)})$ at
$x'$. The fusion condition generates a double pole from
contracting $\bar\Phi_1'^{(ij)}$ with $\Phi_1'^{(kl)}$. Since
$\Phi_1'^{(ij)} = \psi^{(i)}\otimes\psi^{(j)}$, the contraction
$\langle\bar\Phi_1'^{(ij)}\Phi_1'^{(kl)}\rangle\ne 0$ requires
$i=k$ and $j=l$ simultaneously (the two species indices must both
match). Therefore, the terms in the expansion have the form
\begin{widetext}
\begin{eqnarray}
T_{++}^{(ij)}(x)\,T_{++}^{(kl)}(x')
  \sim \frac{c_1/2}{(x-x')^4}\,\delta^{ik}\delta^{jl}
  + \frac{2\,T_{++}^{(ij)}(x')}{(x-x')^2}\,\delta^{ik}\delta^{jl}
  + \cdots \, , \label{eq:TT_OPE}
\end{eqnarray}
\end{widetext}
where $c_1$ is the central charge. The
coefficient $c_1$ is determined by the number of independent
complex components of $\Phi_1'^{(ij)}$. Since
$\Phi_1'^{(ij)}=\psi^{(i)}\otimes\psi^{(j)}$ and each $\psi^{(n)}$
has $d_\mathrm{D}=2$ Dirac polarisations in $(1+1)$d,
\begin{eqnarray}
c_1 = d_\mathrm{D} = 2 \, . \label{eq:c1}
\end{eqnarray}
Summing the OPE~(\ref{eq:TT_OPE}) over all $N^2$ pairs $(i,j)$ for
the total stress tensor eq.~(\ref{eq:Tpp_total}),
\begin{eqnarray}
T_{++}(x)\,T_{++}(x')
  \sim \frac{c/2}{(x-x')^4}
  + \frac{2\,T_{++}(x')}{(x-x')^2}
  + \cdots \, ,
\end{eqnarray}
the leading singularity receives one contribution from each of the
$N^2$ pairs $(i,j)$, giving total central charge
\begin{eqnarray}
c = N^2 \cdot c_1 = 2N^2 \, . \label{eq:c_from_fusion}
\end{eqnarray}
The factor $N^2$ counts the $N^2$ independent composite degrees of
freedom in the $U(N)$ adjoint: $N$ diagonal pairs $(n,n)$ and
$N(N-1)$ off-diagonal pairs $(i,j)$ with $i\ne j$. Had we used
only the diagonal composites, we would have obtained
$c = N \cdot c_1 = 2N$ (the correct result for a $U(N)$ vector
model with one stress tensor per species, but not for the $U(N)$
matrix model in which all $N^2$ adjoint fields are dynamical).

Mode-expanding the total $T_{++}$ along the boundary circle
$x\in[0,2\pi\ell_\mathrm{AdS})$,
\begin{eqnarray}
L_m \equiv \frac{\ell_\mathrm{AdS}}{2\pi}
  \int_0^{2\pi\ell_\mathrm{AdS}} dx\;
  e^{-imx/\ell_\mathrm{AdS}}\,T_{++}(x) \, , \label{eq:Lndef}
\end{eqnarray}
the OPE~(\ref{eq:TT_OPE}) translates into the Virasoro algebra
\begin{eqnarray}
\hspace{-1pc} \left[L_m,\,L_n\right]
  = (m-n)\,L_{m+n}
  + \frac{c}{12}\,m(m^2-1)\,\delta_{m+n,0} \, , \label{eq:Virasoro}
\end{eqnarray}
with central charge $c=2N^2$. This
matches the Brown-Henneaux result $c=3\ell_\mathrm{AdS}/2G_3\sim N^2$
derived in Section~\ref{sec:symmetry}, providing a closed sequence: GN fusion algebra $\to$ $U(N)$ adjoint stress
tensor $\to$ Virasoro algebra $\to$ central charge $c=2N^2$ $\to$
BTZ entropy via Cardy formula. The zero mode $L_0$ has a direct composite-field interpretation.
$L_0$ is the total kinetic energy of all $N^2$ composite fields
$\Phi_1'^{(ij)}$ circulating around the AdS$_3$ boundary, weighted
by their radial wave function $\alpha/z$. In the thermodynamic
limit $L_0 \sim \ell_\mathrm{AdS}^2 M/(16G_3)$.

\section{Emergent AdS$_3$/CFT$_2$}  \label{sec:AdS}

In this section we develop our mapping from the higher-spin GN model to AdS$_3$/CFT$_2$ into a precise duality. Three length scales are evident from the quantities $\lambda k$, $m$, and $|\Delta_1| \sim \Delta_1$. These are, respectively, the momentum scale for the $\Delta_1 \leftrightarrow \Delta_0$ phase transition, the fermion mass, and the spin-1 condensate amplitude. We identify these as the AdS radius $\ell_\mathrm{AdS}= (\lambda k \Delta_1^{1/2})^{-1}$, the Planck length $\ell_\mathrm{P} =  m^{-1}$, and the string length $\ell_\mathrm{S} =( m^{2/3} g  \, \Delta_1^{1/6})^{-1}$. The reason for these identifications will become clear as we establish the holographic mapping. Revisiting some of what we have discussed so far, eq.~(\ref{AdS1}) describes a massive interacting second-rank tensor field propagating on an AdS$_3$ background space-time. The AdS boundary here corresponds to the limit $z^2/\alpha^2 \to 0$ with the horizon limit given by $z^2/\alpha^2 \to \infty$. With the large $N$ scaling $\psi \sim N$, $g^2 \sim 1/N$, $\Phi_1 \sim N^2$, $m \sim N^2$, $\Delta_1 \sim N^4$, which from eq.~(\ref{AdS1}) yields the scaling of the AdS radius $\ell_\mathrm{AdS}  \sim N^2 \alpha = N^2 (\lambda k   \Delta_1^{1/2})^{-1}$, we can read off the AdS curvature
\begin{eqnarray}
R  = \frac{-6}{   \alpha^2 N^4} =     \frac{-6 \lambda^2 k^2  \Delta_1}{   N^4}        \, .
\end{eqnarray}
 In terms of the interactions, we also find that
\begin{eqnarray}
\ell_\mathrm{AdS} =     (g_1'^2 N^{12} )^{1/6}   \frac{ 1 }{m^{2/3} g  \, \Delta_1^{1/6}}    =     (g_1'^2 N^{12} )^{1/6}  \, \ell_\mathrm{S}
\end{eqnarray}
with the effective spin-1 coupling given by
\begin{eqnarray}
g_1'^2   =  \ell_\mathrm{AdS}^6  \,   m^4  g^6  \Delta_1 \sim N^9 \, .
\end{eqnarray}
Note that the AdS radius exceeds the string length by a factor $(g_1'^2 N^{12})^{1/6} \sim N^{7/2}$ at large $N$, consistent with the scale ordering $\ell_\mathrm{S} \ll \ell_\mathrm{P} \ll \ell_\mathrm{AdS}$ of Regime~1.

\subsection{Dual holographic frames}
\label{sec:dualframes}

Our model allows for two dual formulations of AdS/CFT. To see this consider that the emergent coordinate $z = m^{-1}(\Delta_1/\Delta_0^2-1)^{1/2}$ is
real and positive when $\Delta_1 > \Delta_0^2$, i.e.\ on the
spin-1 dominated side of the phase transition (assuming the small phase limit $\Delta_1 \sim |\Delta_1 |$).  Here and throughout,
$\Delta_1 = \langle\bar\Phi_1\Phi_1\rangle$ is the spin-1
\emph{pairing field}, a bound state density measuring the strength
of spin-1 pair correlations, rather than necessarily a true condensate in a single field
$\langle\Phi_1\rangle\neq 0$, which is forbidden in (1+1)d by the Coleman-Mermin-Wagner theorem (see Section~\ref{sec:MerminWagner}). 
 The phase label ``spin-1 condensate phase'' is used throughout as a shorthand
for the region $\Delta_1 > \Delta_0^2$ where the spin-1 pairing field
dominates; true long-range order $\langle\Phi_1\rangle\neq 0$ is
recovered only in the large-$N$ dual description via the
species-sum regulation mechanism which we will discuss in Section~\ref{sec:MerminWagner}.  On the opposite side of the transition,
$\Delta_0^2 > \Delta_1$, the natural emergent coordinate is instead
\begin{eqnarray}
\tilde{z} \equiv \frac{1}{m}\left(\frac{\Delta_0^2}{\Delta_1}\right)^{1/2} ,
\label{eq:ztilde}
\end{eqnarray}
which is real and positive precisely when $z^2 < 0$, i.e.\ when the
$z$ description breaks down.  The two coordinates together cover the
full GN phase diagram. Frame~1 (coordinate $z$) covers the spin-1
dominated phase $\Delta_1 > \Delta_0^2$, and Frame~2 (coordinate
$\tilde{z}$) covers the spin-0 dominated phase $\Delta_0^2 >
\Delta_1$.  Crucially, the phase transition point $\Delta_1 =
\Delta_0^2$ maps to $z = 0$ in Frame~1 (the AdS boundary) but to
$\tilde{z} = m^{-1} \equiv \ell_\mathrm{P}$ in Frame~2, a
\emph{finite} location inside the bulk.  The two frames are related
by the $\mathbb{Z}_2$ symmetry $\Delta_1/\Delta_0^2 \leftrightarrow
\Delta_0^2/\Delta_1$ of the GN phase diagram, which exchanges the
roles of the two condensates. The following is a detailed characterization of the two frames:

\paragraph{Frame 1: spin-1 condensate phase and Vasiliev higher-spin holography}

The coordinate $z$ satisfies $z\to 0$ when $\Delta_1\to\Delta_0^2$
(the phase transition) and $z\to\infty$ when $\Delta_1\gg\Delta_0^2$
(deep spin-1 condensate). In the Poincar\'e patch metric
$ds^2 = (\alpha^2/z^2)(-dt^2+dx^2+dz^2)$ the boundary $z\to 0$ is
therefore the chiral critical point: both condensates are equal,
the theory sits at its strongly coupled conformal fixed point, the
open string masses $M_{ij} = \kappa^{1/2}|z^{(i)}-z^{(j)}|$ vanish,
and the entire higher-spin tower $\{\Phi_s\}$ becomes massless and
degenerate. This is the Vasiliev point~\cite{Vasiliev1990,VasilievReview} of
an infinite tower of massless higher-spin gauge fields dual to a
strongly coupled boundary CFT. The deep bulk $z\to\infty$ is
instead the ordered spin-1 phase, where the composites propagate as
nearly free massive particles and the geometry is classical
(Regime~1, $\ell_\mathrm{S}\sim\ell_\mathrm{P}\ll\ell_\mathrm{AdS}$).
Frame~1 therefore has an \emph{inverted} classical/quantum ordering relative to conventional holography. Here, specifically, the boundary is the
most strongly coupled and nonclassical region, and the bulk is
tractable precisely because moving away from the boundary means
moving away from the critical point into the ordered spin-1 phase.

\paragraph{Frame 2: spin-0 condensate phase and conventional AdS/CFT}

In the spin-0 dominated phase $\Delta_0^2 > \Delta_1$, the material
derivative for $\tilde{z}$ is
\begin{eqnarray}
\partial_\mu\tilde{z}
  = -\frac{\tilde{z}}{2\Delta_1}\,\partial_\mu\Delta_1 \, ,
\label{eq:dztilde}
\end{eqnarray}
which weights fluctuations of $\Delta_1$ by the local condensate
amplitude.  Retaining leading-order terms, the kinetic term in the
$\tilde{z}$ frame takes the AdS$_3$ Poincar\'e form
\begin{eqnarray}
ds^2 = \frac{\tilde\alpha^2}{\tilde{z}^2}
       \left(-dt^2+dx^2+d\tilde{z}^2\right) ,
\label{eq:metrictilde}
\end{eqnarray}
where $\tilde\alpha = (\tilde\lambda\tilde k\,\Delta_0)^{-1}$ is the
AdS radius in Frame~2, set by the spin-0 condensate rather than the
spin-1 condensate. The hierarchy is now the conventional one. The
boundary $\tilde{z}\to 0$ has $\Delta_1\gg\Delta_0^2$: the
composite fields are massive and well-defined and the geometry is
maximally classical. The phase transition $\Delta_1=\Delta_0^2$ resides at the \emph{finite} bulk depth
$\tilde{z} = \ell_\mathrm{P} = m^{-1}$, where the geometry turns
from classical to nonclassical, and the deep bulk
$\tilde{z}\to\infty$ ($\Delta_1\to 0$) is the far interior of the
chirally broken phase, highly nonclassical and stringy. Frame~2
thus realizes the standard holographic picture and is the natural
language for conventional AdS/CFT.

In Frame~2 the scalar condensate $\Delta_0$ plays the standard
holographic role of a relevant scalar deformation, which is negligible near
the boundary ($\Delta_0\to 0$, chiral symmetry restored), growing
toward the interior as $\Delta_0\sim\tilde{z}^{-1}$, and reaching
full strength at the Planck depth $\tilde{z} = \ell_\mathrm{P}$
(chiral symmetry fully broken, massive fermions). This is
precisely the conventional AdS/CFT dictionary for a relevant
operator sourcing RG flow from a UV fixed point into a strongly
coupled IR phase (the bulk field dual to a scalar at the
Breitenlohner-Freedman bound~\cite{BF1982}).

To obtain the Lagrangian for Frame 2, we substitute $\Delta_1 = \Delta_0^2/(m^2\tilde{z}^2)$ into the
spin-1 Lagrangian and introduce the rescaled field
$\tilde\Phi_1' \equiv (\tilde\lambda\tilde k\,\Delta_0^{1/2}
m^{-3/2})\Phi_1$. The composite-field Lagrangian then takes the
AdS$_3$ form
\begin{eqnarray}
\widetilde{\mathcal{L}}_{\Phi_1}'
  &=& \frac{\tilde\alpha}{\tilde{z}}
      \left(\partial_{\mu}\bar{\tilde\Phi}_1'\partial^{\mu}\tilde\Phi_1'
           +\partial_{\tilde{z}}\bar{\tilde\Phi}_1'
            \partial_{\tilde{z}}\tilde\Phi_1'\right) \,\nonumber  \\
             &-&  \frac{\tilde{z}}{\tilde\alpha}
      \left[\tilde{m}_1'^2\,\bar{\tilde\Phi}_1'\tilde\Phi_1'
      +\frac{\tilde{g}_1'^2}{2}
       \left(\bar{\tilde\Phi}_1'\tilde\Phi_1'\right)^2\right] ,
\label{eq:AdS1tilde}
\end{eqnarray}
where the rescaled mass and coupling are
\begin{eqnarray}
\hspace{-1pc} \tilde{m}_1'^2
  = \frac{m^5 g^2\tilde{z}^2}{(\tilde\lambda\tilde k)^4\Delta_0^2}
      \left(\frac{g^4\Delta_0^2}{m^4\tilde{z}^2} - 1\right) , \hspace{1pc} \tilde{g}_1'^2
  =  \frac{m^8 g^6\tilde{z}^6}{(\tilde\lambda\tilde k)^6\Delta_0^6} \, .
\label{eq:gtilde}
\end{eqnarray}
The kinetic structure of $\widetilde{\mathcal{L}}_{\Phi_1}'$ is
identical to the standard AdS$_3$ form established in
Section~\ref{sec:measure}, eq.~(\ref{eq:Sbulkstandard}), with
$\tilde{z}$ and $\tilde\alpha$ replacing $z$ and $\alpha$.  The
effective mass $\tilde{m}_1'^2$ vanishes at the boundary ($\tilde{z}\to 0$,
$\Delta_1\to\infty$) and diverges at the critical point
$\tilde{z}=\ell_\mathrm{P}$, signaling the breakdown of the
spin-1 description at the phase transition, where Frame~1
takes over.

\begin{table*}[t!]
\renewcommand{\arraystretch}{1.5}
\begin{tabular}{lll}
\hline\hline
 & Frame 1 ($z$ coordinate) & Frame 2 ($\tilde{z}$ coordinate) \\
\hline
Domain
  & $\Delta_1 > \Delta_0^2$
  & $\Delta_0^2 > \Delta_1$ \\
Radial coord.
  & $z = m^{-1}(\Delta_1/\Delta_0^2-1)^{1/2}$
  & $\tilde{z} = m^{-1}(\Delta_0^2/\Delta_1)^{1/2}$ \\
Boundary ($z,\tilde{z}\to 0$)
  & critical point $\Delta_1=\Delta_0^2$, nonclassical
  & $\Delta_1\to\infty$, $\Delta_0\to 0$, maximally classical \\
Critical point
  & boundary ($z=0$)
  & finite bulk depth ($\tilde{z}=\ell_\mathrm{P}$) \\
Deep bulk ($z,\tilde{z}\to\infty$)
  & $\Delta_1\gg\Delta_0^2$, classical gravity
  & $\Delta_1\to 0$, $\Delta_0$ fully ordered, stringy \\
Classical/quantum
  & boundary nonclassical $\to$ bulk classical
  & boundary classical $\to$ bulk nonclassical \\
Strings emerge
  & near boundary
  & deep in bulk, beyond $\tilde{z}=\ell_\mathrm{P}$ \\
Holographic analogue
  & Vasiliev higher-spin holography
  & conventional AdS/CFT \\
\hline\hline
\end{tabular}
\caption{Comparison of the two holographic frames admitted by the
emergent AdS$_3$ geometry.  The frames cover opposite sides of the
GN phase transition and are related by the $\mathbb{Z}_2$ symmetry
$\Delta_1/\Delta_0^2\leftrightarrow\Delta_0^2/\Delta_1$.  The phase transition sits on the
boundary in Frame~1 and at finite bulk depth $\tilde{z}=\ell_\mathrm{P}$
in Frame~2.  ``Classical'' refers throughout to the spin-1
pairing field geometry ($\Delta_1$ sector); $\Delta_0$ plays the role
of the boundary CFT source in both frames.}
\label{tab:frames}
\end{table*}

\subsection{The two frames and $\mathbb{Z}_2$ symmetry}

The transformation $\Delta_1/\Delta_0^2 \leftrightarrow
\Delta_0^2/\Delta_1$ is a $\mathbb{Z}_2$ symmetry of the GN phase
diagram that exchanges the two condensates.  Under this symmetry
$z \leftrightarrow i\tilde{z}$ (up to the factor of $m$), reflecting
the fact that the two coordinates are defined on opposite sides of
the phase transition.  The explicit relation between $z$ and
$\tilde{z}$ follows from substituting $\Delta_1 =
\Delta_0^2/(m^2\tilde{z}^2)$ into $z^2 = m^{-2}(\Delta_1/\Delta_0^2-1)$:
\begin{eqnarray}
z^2 = \frac{1}{m^2}\left(\frac{1}{m^2\tilde{z}^2} - 1\right)
    = \frac{1 - m^2\tilde{z}^2}{m^4\tilde{z}^2} \, .
\label{eq:zztilde}
\end{eqnarray}
This confirms that $z=0$ (Frame~1 boundary, critical point) corresponds
to $\tilde{z}=m^{-1}=\ell_\mathrm{P}$ (finite bulk depth in Frame~2);
$z>0$ (Frame~1 physical region) requires $\tilde{z}<\ell_\mathrm{P}$
(the classical region of Frame~2 between the boundary and the
critical point); and $z^2<0$ (the unphysical region of Frame~1)
corresponds to $\tilde{z}>\ell_\mathrm{P}$ (Frame~2 beyond the
critical point, into the strongly coupled interior).

We shall see that this $\mathbb{Z}_2$ exchange is the holographic counterpart of the
open/closed string T-duality derived in future work~\cite{CompanionStrings}, now acting on the \emph{pairing field ratio} rather than the angular radius.
Just as T-duality exchanges winding and momentum modes at the
self-dual radius $R_\theta=\ell_\mathrm{S}$, the pairing field exchange
$\Delta_1/\Delta_0^2\leftrightarrow\Delta_0^2/\Delta_1$ maps one
holographic frame to the other at the self-dual point
$\Delta_1=\Delta_0^2$, i.e.\ the phase transition.
The key properties of both frames are summarized in Table~\ref{tab:frames}.

\subsection{Three length scales and bulk regimes}
\label{sec:regimes}

The analysis in this subsection is carried out in Frame~2, where the
boundary geometry is classical and the three length scales have a
transparent physical interpretation in terms of the conventional
AdS/CFT hierarchy.  From our earlier results, we see that finite $N$ and large $k$ send the theory towards the
horizon where local curvatures are large. In contrast, the boundary limit
corresponds to large $N$, small $k$ and small curvature.  The
three length scales $\ell_\mathrm{AdS}$, $\ell_\mathrm{P}$, and
$\ell_\mathrm{S}$ define three distinct holographic regimes,
summarized in Table~\ref{table1}, separated by the Hawking-Page
transition at $T_\mathrm{HP}\sim\ell_\mathrm{AdS}^{-1}$ and the
Hagedorn transition at $T_H\sim\ell_\mathrm{S}^{-1}$. We will provide a full
dictionary of all critical scales and temperatures and their
meanings in the GN model, AdS bulk, and string/higher-spin
language in future work~\cite{CompanionStrings}.  The precise
field-theory conditions for each regime are as follows.

  \begin{enumerate}

  \item   $\ell_\mathrm{S} \sim  \ell_\mathrm{P}   \ll  \ell_\mathrm{AdS}$. This ordering of scales corresponds to the near-boundary regime described by classical gravity where strings are quantum mechanical. On the field theory side, this is equivalent to $g_1'^2 N^{12} \gg 1$  and     $1     \sim    \Delta_1 /(g^3  \Delta_0^2)    \ll         g^{14}  N^{12} \Delta_0^4  /(\lambda k \Delta_1^{1/2})^6 $, satisfied near criticality ($g \sim 1$, $\Delta_1/\Delta_0^2 \gtrsim 1$) at large $N$ and small $k$, with large spin-1 effective coupling: a strongly correlated conformal field theory in asymptotically flat $(1+1)$-dimensional space-time. The correspondence is
  \begin{eqnarray}
\hspace{2pc}   \emph{Classical $AdS_3$}  \;  \leftrightarrow  \;  \emph{ Strongly Correlated $CFT_2$  }  \nonumber
  \end{eqnarray}

  \item   $\ell_\mathrm{S}  <    \ell_\mathrm{P}   \sim   \ell_\mathrm{AdS}$. This is the intermediate bulk region described by quantum gravity, equivalent in the field theory to the conditions $g_1'^2 N^{12} \sim 1$  and     $1     \lesssim     \Delta_1 /(g^3  \Delta_0^2)    \sim         g^{14}  N^{12} \Delta_0^4  /(\lambda k \Delta_1^{1/2})^6 $, which is satisfied when $g \gtrsim 1$, $\Delta_1/\Delta_0^2 \gtrsim 1$, and $N^{12}/(\lambda k)^6 \sim 1$. The correspondence in this region reads
  \begin{eqnarray}
\hspace{1pc}   \emph{Quantum Gravity}  \;  \leftrightarrow  \;  \emph{Gauge Field Theory  }  \nonumber
  \end{eqnarray}

  \item      $\ell_\mathrm{S} \sim  \ell_\mathrm{AdS}   \ll   \ell_\mathrm{P}$. This is the deep interior of the bulk, best described by a quantum theory of strings, with field theory conditions $g_1'^2 N^{12} \ll  1$  and     $1     \   \sim         g^{14}  N^{12} \Delta_0^4  /(\lambda k \Delta_1^{1/2})^6  \ll   \Delta_1 /(g^3  \Delta_0^2)$, satisfied when $g > 1$, $\Delta_1/\Delta_0^2 \gg  1$, $N^{12}/(\lambda k)^6 \ll 1$. The spin-1 field is classical with large mean-field pairing, but $\ell_\mathrm{P} \gg \ell_\mathrm{AdS}$ places the Planck scale far below the AdS scale, so quantum string effects dominate over classical gravity while the fermions are strongly correlated with a small scalar condensate and mass. The correspondence here is
  \begin{eqnarray}
\hspace{2pc}   \emph{String Field Theory}  \;  \leftrightarrow  \;  \emph{Classical Gauge Theory}  \nonumber
  \end{eqnarray}

  \end{enumerate}

\begin{table*}[t!]
\centering
{
\renewcommand{\arraystretch}{1.5}
\begin{tabular}{p{3.2cm} p{3.8cm} p{3.8cm} p{3.8cm}}
\hline\hline
 & Regime 1 & Regime 2 & Regime 3 \\
\hline
Scale hierarchy
  & $\ell_\mathrm{S}\sim\ell_\mathrm{P}\ll\ell_\mathrm{AdS}$
  & $\ell_\mathrm{S}<\ell_\mathrm{P}\sim\ell_\mathrm{AdS}$
  & $\ell_\mathrm{S}\sim\ell_\mathrm{AdS}\ll\ell_\mathrm{P}$ \\
Bulk description
  & Classical AdS$_3$ gravity
  & Quantum gravity
  & Quantum string theory \\
Boundary description
  & Strongly correlated CFT$_2$
  & Gauge field theory
  & Classical gauge theory \\
GN parameters
  & $N\to\infty$, large $\lambda$
  & $N\to\infty$, any $\lambda$
  & Any $N$ and $\lambda$ \\
Condensate ratio
  & $\Delta_1\gg\Delta_0^2$
  & $\Delta_1\gtrsim\Delta_0^2$
  & $\Delta_1\sim\Delta_0^2$ \\
Holographic frame
  & Conventional AdS/CFT
  & Transition region
  & Vasiliev higher-spin \\
String coupling
  & $g_s\to 0$,\; $\ell_\mathrm{S}^2/\ell_\mathrm{AdS}^2\to 0$
  & $g_s\to 0$,\; $\ell_\mathrm{S}^2/\ell_\mathrm{AdS}^2\neq 0$
  & $g_s\neq 0$,\; $\ell_\mathrm{S}^2/\ell_\mathrm{AdS}^2\neq 0$ \\
Temperature
  & $T < T_\mathrm{HP}$
  & $T_\mathrm{HP} < T < T_H$
  & $T > T_H$ \\
\hline\hline
\end{tabular}
}
\caption{Three holographic regimes of the emergent AdS$_3$/CFT$_2$
correspondence derived from the GN model, ordered from weakest (left)
to strongest (right) bulk quantum effects, separated by the Hawking-Page
transition at $T_\mathrm{HP}\sim\ell_\mathrm{AdS}^{-1}$ and the Hagedorn
transition at $T_H\sim\ell_\mathrm{S}^{-1}$.~\cite{Maldacena,GiveonKutasovSeiberg}}
\label{table1}
\end{table*}

\subsection{Symmetry matching and the holographic dictionary}
\label{sec:symmetry}

We now address the key symmetries that link the GN model with the
emergent bulk. The $\mathrm{AdS}_3$ isometry group
$SO(2,2)\cong SL(2,\mathbb{R})\times SL(2,\mathbb{R})$ constrains
which GN composite fields can appear at which radial depth. First, we note that the
spin-0 condensate $\Delta_0=\langle\bar\psi\psi\rangle$ transforms
as a bulk scalar with mass $m^2\ell_\mathrm{AdS}^2 = -1$, sitting
precisely at the Breitenlohner-Freedman bound~\cite{BF1982}. It is
dual to the boundary operator $\bar\psi\psi$ of conformal dimension
$\Delta=1$, the relevant deformation that drives the GN chiral
transition. Second, that the spin-1 composite $\Phi_1$ is a rank-2 tensor whose
vector component $\varphi^\mu$ (Section~\ref{sec:EH}) is dual to the
conserved current $J^\mu = \bar\psi\gamma^\mu\psi$ of dimension
$\Delta=2$. Its emergent $U(1)$ gauge symmetry is
the bulk manifestation of this conservation law. The higher-spin
composites $\Phi_s$ are dual to the spin-$s$ currents
$J^{\mu_1\cdots\mu_s}\sim\bar\psi\partial^{s-1}\psi$ of dimension
$\Delta=s+1$, with bulk masses $m_s^2\ell_\mathrm{AdS}^2 = s^2-1$,
consistent with the linear Regge trajectory $M_s^2\propto s^2$.

The conformal dimensions above follow from the standard
AdS$_3$/CFT$_2$ mass-dimension relation for a boundary theory with
$d=2$ spacetime dimensions,
\begin{eqnarray}
\Delta(\Delta - 2) = m^2\ell_\mathrm{AdS}^2 \, ,
\label{eq:massdim}
\end{eqnarray}
which gives $\Delta = 1 \pm \sqrt{1 + m^2\ell_\mathrm{AdS}^2}$. For
the scalar condensate $\Delta_0$ with $m^2\ell_\mathrm{AdS}^2 = -1$,
the two roots are $\Delta_- = 0$ and $\Delta_+ = 2$. The
normalizable mode $\Delta_+=2$ is the response and the
non-normalizable mode $\Delta_-=0$ is the source. Since this
operator sits precisely at the BF bound, one may alternatively
quantize using $\Delta_-=0$ as the response~\cite{BF1982}, in which
case the source dimension is $\Delta=1$ for the relevant deformation
$\bar\psi\psi$. For the higher-spin bulk fields with
$m_s^2\ell_\mathrm{AdS}^2 = s^2-1$, eq.~(\ref{eq:massdim}) gives
\begin{eqnarray}
\Delta = 1 + \sqrt{1 + s^2 - 1} = 1 + s \, ,
\label{eq:deltaS}
\end{eqnarray}
reproducing the conformal dimensions $\Delta=2, 3, \ldots, s+1$ of
the spin-$s$ conserved currents $J^{\mu_1\cdots\mu_s}$ directly from
the Regge mass formula. The higher-spin tower is therefore
fixed by a single formula, the Regge trajectory, with each level
matching a conserved current of the GN boundary theory.

The reader may notice that the isometry group
$SO(2,2)\cong SL(2,\mathbb{R})\times SL(2,\mathbb{R})$ of the bulk is
not the full symmetry of the boundary theory. In AdS$_3$, the
asymptotic symmetry group is enhanced from
$SL(2,\mathbb{R})\times SL(2,\mathbb{R})$ to two copies of the
Virasoro algebra by the Brown-Henneaux
mechanism~\cite{BrownHenneaux}. Diffeomorphisms that preserve the
AdS$_3$ boundary conditions but are not globally well-defined
generate an infinite-dimensional asymptotic symmetry
$\mathrm{Diff}(S^1)\times\mathrm{Diff}(S^1)$, with central charge
\begin{eqnarray}
c = \frac{3\ell_\mathrm{AdS}}{2G_3}
  = 6\pi\ell_\mathrm{AdS} \sim N^2 \, ,
\label{eq:BHcentral}
\end{eqnarray}
using $G_3 = 1/4\pi$. This is not an independent input. It is
precisely the Virasoro algebra derived in
Section~\ref{sec:higherspin} from the fusion algebra of the GN
composites in eq.~(\ref{eq:Virasoro}). The symmetry matching therefore
closes through 
\begin{widetext}
\begin{eqnarray}
\underbrace{\text{GN fusion algebra}}_{\text{boundary, microscopic}}
\;\longrightarrow\;
\underbrace{\mathrm{Vir}\times\mathrm{Vir}}_{\text{boundary, emergent}}
\;\longleftrightarrow\;
\underbrace{\mathrm{Diff}(S^1)\times\mathrm{Diff}(S^1)}_{\text{bulk, asymptotic}}
\, . \nonumber
\end{eqnarray}
\end{widetext}
The global $SL(2,\mathbb{R})\times SL(2,\mathbb{R})$ sub-algebra of
the Virasoro algebra matches the AdS$_3$ isometry group. The full
Virasoro algebra captures the complete tower of boundary Ward
identities satisfied by the GN composite operators, which is the
constraint that each $\Phi_s$ sits at the correct conformal
dimension $\Delta=s+1$ on the boundary.

\subsection{The AdS$_3$ isometries in condensate language}
Here, we give a more explicit rendering of symmetries in terms of the underlying condensed matter structures. We begin by considering that the isometry group $SO(2,2)$ of the emergent AdS$_3$ has six
generators, and each of these can be interpreted in terms of GN
condensate variables $(t, x, z)$, where
$z = m^{-1}(\Delta_1/\Delta_0^2 - 1)^{1/2}$. The two
\emph{boundary translations} $P_\mu$ and the \emph{boundary
Lorentz boost} $M_{01}$, acting at fixed $z$, are the expected
Poincar\'e symmetries of the $(1+1)$d GN theory at fixed
condensate ratio. The \emph{dilatation} $D$,
$(x^\mu, z) \to \lambda(x^\mu, z)$, rescales both boundary
coordinates and radial depth simultaneously. In GN language this is
the renormalisation group transformation: rescaling the energy
scale and the condensate ratio together, with
$\Delta_1/\Delta_0^2 \to \lambda^2 \Delta_1/\Delta_0^2$ shifting $z$
proportionally. This is the bulk manifestation of scale invariance
at the chiral fixed point $\Delta_1 = \Delta_0^2$ (the AdS
boundary).

The two \emph{special conformal transformations} $K_\mu$ are the
most physically significant. In Poincar\'e coordinates they act as
$x^\mu \to x^\mu + b^\mu z^2 + \cdots$,
$z \to z + b_\mu x^\mu z + \cdots$, mixing the boundary directions
with the radial direction. In GN language, $K_\mu$ states that a
boundary translation that depends on the condensate depth (a
$z$-dependent shift of $x^\mu$) is equivalent to a change of the
condensate ratio that depends on boundary position (an
$x^\mu$-dependent shift of $z$). This is precisely the
density-phase coupling $(\partial_\mu z)\partial_z$ of the material
derivative in which a local fluctuation of $\Delta_1/\Delta_0^2$ at a given
boundary point is equivalent to a boundary translation at a given
condensate depth. The special conformal transformations are
therefore the material derivative written as a \emph{symmetry}
rather than a dynamical equation. The fact that $K_\mu$ closes into
the conformal algebra
$[K_\mu, P_\nu] = 2(\eta_{\mu\nu}D - M_{\mu\nu})$ with $P_\mu$ and
$D$ states that the density-phase coupling is consistent with
boundary Poincar\'e invariance and scale invariance. This is a
non-trivial constraint that the GN model satisfies because the
condensate ratio $\Delta_1/\Delta_0^2$ transforms correctly under
the chiral RG flow.

The key feature of Frame~1 is the inverse relationship between the
spin-0 and spin-1 condensates encoded in $z = m^{-1}(\Delta_1/\Delta_0^2-1)^{1/2}$. The AdS$_3$ bulk exists
only for $\Delta_1>\Delta_0^2$. This combination is natural from
the perspective of the AdS$_3$ dilatation isometry
$z\to\lambda z$, $x^\mu\to\lambda x^\mu$. Near the boundary the
condensates scale as $\Delta_0\sim z^{\Delta_-}= z^1$ and
$\Delta_1\sim z^{\Delta_+}=z^2$ in accordance with their conformal
dimensions, so the ratio $\Delta_1/\Delta_0^2\sim z^0$ is
\emph{neutral} under dilatations. This makes it a good radial
coordinate, insensitive to the overall conformal rescaling that translates along the boundary, and capturing only the genuinely
radial degree of freedom.

The geometric content of AdS$_3$ becomes particularly
transparent in Frame~1 via the polar decomposition
$\Phi_1' = \rho\,e^{i\theta}$, where $\rho = |\Phi_1'|$ is the
condensate amplitude and $\theta$ is the Goldstone phase of the
spontaneously broken $U(1)$ symmetry $\Phi_1'\to e^{i\alpha}\Phi_1'$. The kinetic term separates to give 
\begin{eqnarray}
\frac{\alpha}{z}\,\partial_A\bar\Phi_1'\partial^A\Phi_1'
  = \frac{\alpha}{z}\left(
      \partial_A\rho\,\partial^A\rho
    + \rho^2\,\partial_A\theta\,\partial^A\theta
    \right) , \label{eq:polar}
\end{eqnarray}
with the two sectors having distinct geometric roles. The amplitude
$\rho$ maps to the radial direction. Since $\Delta_1\sim\rho^2$,
the coordinate is $z\propto\rho/\Delta_0$, so varying $\rho$ at
fixed $\Delta_0$ is varying depth, and the equation of motion for
$\rho$ is a radial wave equation on AdS$_3$. The phase $\theta$, a compact scalar with
periodicity $2\pi$, maps to the angular direction of global
AdS$_3$; in the condensed phase, with $\rho$ frozen at
$\rho_0^2 = m_1'^2/g_1'^2$, its kinetic term
$(\alpha\rho_0^2/z)\partial_A\theta\,\partial^A\theta$ describes a
compact scalar whose winding modes are the closed string winding
modes which we discuss in our follow up paper~\cite{CompanionStrings}. The Mexican hat potential
$-m_1'^2\rho^2+(g_1'^2/2)\rho^4$ has its minimum at $\rho_0$,
corresponding to a preferred radial position
$z_0\sim\rho_0/\Delta_0$, or equivalently
$\tilde{z}_0 < \ell_\mathrm{P}$ in Frame~2.

This picture is consistent with the string description of~\cite{CompanionStrings} emerging near the Frame~1 boundary
rather than in the deep interior. Specifically, as $z\to 0$ the higher-spin
tower degenerates at the chiral transition analogous to the tensionless limit of Vasiliev
theory~\cite{VasilievReview}, precisely where the boundary CFT is
most strongly coupled and
$\ell_\mathrm{S}\sim\ell_\mathrm{AdS}$. The classical bulk gravity
regime $\ell_\mathrm{P}\ll\ell_\mathrm{S}\ll\ell_\mathrm{AdS}$ is
realized at intermediate $z$, where the spin-1 condensate
dominates and the composites propagate as nearly free massive
particles: the tractable weakly interacting regime identified in
the introduction.

Finally, we note that the three-dimensional Newton constant $G_3$ is
not an independent input. It is fixed by requiring that the
emergent geometry reproduce the Brown-Henneaux central charge
$c = 3\ell_\mathrm{AdS}/2G_3$ with $c \sim N^2$ from the Virasoro
calculation eq.~(\ref{eq:c_from_fusion}). This gives
\begin{eqnarray}
G_3 = \frac{3\ell_\mathrm{AdS}}{2c} \sim \frac{3\ell_\mathrm{AdS}}{2N^2 c_1} \, ,
\label{eq:G3fromN}
\end{eqnarray}
where $c_1 = \mathcal{O}(1)$ is the per-species central charge
[eq.~(\ref{eq:c_from_fusion})]. Expressing $\ell_\mathrm{AdS}$ in
terms of microscopic parameters, gives
\begin{eqnarray}
G_3
  \sim \frac{\ell_\mathrm{AdS}}{N^2}
  = \frac{1}{N^2 \lambda k \Delta_1^{1/2}} \, .
\label{eq:G3explicit}
\end{eqnarray}
The ratio $\ell_\mathrm{P}/\ell_\mathrm{AdS} = m/(\lambda k\Delta_1^{1/2})$
is $\mathcal{O}(1/N^2)$ at large $N$ (using $m\sim N^2$,
$\Delta_1\sim N^4$). Using $\ell_\mathrm{AdS}\sim N^2\ell_\mathrm{P}$,
this simplifies to
\begin{eqnarray}
G_3 \sim \frac{\ell_\mathrm{P}}{N^2} \, ,
\label{eq:G3scaling}
\end{eqnarray}
consistent with the standard relation $G_3\sim g_s^2\ell_s/N^2$ for
$N$ D1-branes~\cite{Polchinski1998}. This large-$N$ counting
argument gives the scaling of $G_3$ with the microscopic parameters.
A second, independent derivation from the spin-2 sector of the
composite field Lagrangian $\mathcal{L}_{\Phi_1}'$ gives the exact
value $G_3 = 1/4\pi$ (in units of $\ell_\mathrm{AdS}$), derived in
Section~\ref{sec:EH}.

\section{Mermin-Wagner, Coleman theorems, and the emergence of true condensates}
\label{sec:MerminWagner}

In $(1+1)$ spacetime dimensions the Coleman
theorem~\cite{Coleman1973} and its finite-temperature extension,
the Mermin-Wagner theorem~\cite{MerminWagner1966} (CMW), place
strong restrictions on spontaneous symmetry breaking. The
condensates of the GN model are affected differently according to
their symmetry-breaking patterns. The scalar condensate
$\Delta_0 = \langle\bar\psi\psi\rangle$ breaks only the discrete
$\mathbb{Z}_2$ chiral symmetry $\psi\to\gamma^5\psi$, to which
these theorems do not apply; at large $N$ it develops a genuine
non-zero expectation value at $T=0$, with the broken phase
persisting up to the crossover temperature
$T_P\sim m$~\cite{GrossNeveu}. The spin-1 sector is subtler. For
the diagonal composite
$\Phi_1^{(ii)} = \rho^{(i)}e^{i\theta^{(i)}}$, whose continuous
$U(1)_1$ phase symmetry the theorem does govern, infrared
fluctuations produce a logarithmically divergent propagator
$\langle\theta(x)\theta(0)\rangle\sim\ln|x|$, preventing
$\langle\Phi_1^{(ii)}\rangle\neq 0$ and leaving quasi-long-range
order with algebraically decaying correlations
$\langle\Phi_1^{(ii)}(x)\Phi_1^{(ii)}(0)\rangle\sim|x|^{-\eta(T)}$~\cite{Petrov2000,Petrov2001}.
The pair density $\Delta_1^{(ii)} = \langle|\Phi_1^{(ii)}|^2\rangle$
can nevertheless remain nonzero, since
$\langle|\Phi_1|^2\rangle \geq |\langle\Phi_1\rangle|^2$: it is the
phase coherence, not the pair density, that is obstructed. The
three condensation channels that we are considering are affected by CMW in qualitatively different
ways~\cite{Witten1978}:

\begin{enumerate}

\item The $\Delta_0$ channel:
$\Delta_0 = \frac{1}{N}\sum_n\langle\bar\psi^{(n)}\psi^{(n)}\rangle$.
The species label evades Pauli exclusion (one fermion of each
species occupies the same spatial mode), and the $O(N)$ coherent
sum makes $\Delta_0$ a classical field with relative fluctuations
of order $1/\sqrt{N}$; this is the mechanism Witten
identified~\cite{Witten1978}, with the species label playing the
role that occupation number plays in BEC. CMW does not apply
(discrete symmetry), hence this is a true condensate.

\item The off-diagonal $\Delta_1^{(ij)}$ channel ($i\neq j$):
$\Phi_1^{(ij)} = \psi^{(i)}\otimes\psi^{(j)}$ pairs distinct
species, with no Pauli obstruction and $O(N^2)$ coherent pairs
contributing to the same spatial mode, making it the most
classical object in the model. CMW applies in principle to the
pair's continuous $U(1)$ phase, but the $O(N^2)$ enhancement
strongly suppresses phase fluctuations, characterized by true long-range order.

\item The diagonal $\Delta_1^{(ii)}$ channel:
$\Phi_1^{(ii)} = \psi^{(i)}\otimes\psi^{(i)}$ pairs two fermions
of the same species, with Pauli satisfied by the distinct Dirac
indices ($\alpha\neq\beta$ in $\Phi_{1,\alpha\beta}^{(ii)}$).
Only one pair per species contributes to each spatial mode, the
enhancement is only $O(N)$, and CMW restricts the channel to
quasi-long-range order. This is the channel described by the BKT
analysis of our follow-up paper~\cite{CompanionStrings} where the phase $\theta^{(i)}$ winds around
vortex cores.

\end{enumerate}

From a microscopic viewpoint, local fluctuations of the spin-1 condensate density and phase, mediated by the
material derivative
$\partial_\mu\to\partial_\mu+(\partial_\mu z)\partial_z$, are qualitatively different for the two regimes separated by the Planck length
$\ell_\mathrm{P} = m^{-1}$, the scale at which
$\Delta_1\sim\Delta_0^2$. In the long-wavelength regime
($k\ll\ell_\mathrm{P}^{-1}$, large phase stiffness $\rho_s$, near
the boundary), the amplitude resides at its mean-field value and
phase fluctuations propagate as collective Bogoliubov phonons through
the relay of off-diagonal composites $\Phi_1^{(n,n+1)}$ linking
adjacent D1-branes. Such sequential hops average at large
$N$ into a continuous radial propagation. This ``open string relay'' is simply the bulk name for the standard Bogoliubov phonon channel in the boundary description which functions precisely when the
condensate is coherent enough for the Bogoliubov approximation to
hold; this is the classical gravitational wave regime. In the
short-wavelength regime ($k\gg\ell_\mathrm{P}^{-1}$, small
$\rho_s$, deep bulk), amplitude and phase fluctuate independently,
the relay breaks down, and the signal is carried instead by
discrete closed-string winding modes propagating as individual
quantum events (the quantum foam regime). The two channels coexist
in any condensate, with relative weight controlled by $\rho_s$,
and the crossover at $k\sim\ell_\mathrm{P}^{-1}$ marks the
boundary between Regimes~1 and~2 of Section~\ref{sec:regimes}.

It is important to point out that the large-$N$ species sum of Section~\ref{sec:measure} provides IR
regulation of the phase propagator. Consider that each species $n$ at depth
$z^{(n)}$ contributes a local effective mass $m_\theta^2(z^{(n)})$
set by the local condensate ratio
$\Delta_1^{(n)}/\Delta_0^{(n)2}$. The total dressed phase
propagator is
\begin{eqnarray}
G_\theta(x)
  = \sum_{n=1}^N
    \int\frac{d^2k}{(2\pi)^2}\;
    \frac{e^{ik\cdot x}}{\rho_s k^2 + m_\theta^2(z^{(n)})} \, .
\label{eq:Gthetasum}
\end{eqnarray}
In the large-$N$ limit, with species distributed continuously via
$\rho_*(z)=z/\alpha$, this becomes
\begin{widetext}
\begin{eqnarray}
G_\theta(x)
  \;\xrightarrow{N\to\infty}\;
  \int_0^\infty dz\;\rho_*(z)
    \int\frac{d^2k}{(2\pi)^2}\;
    \frac{e^{ik\cdot x}}{\rho_s k^2 + m_\theta^2(z)}
  \;=\;
  \frac{1}{|x|} \, ,
\label{eq:Gthetadressed}
\end{eqnarray}
\end{widetext}
the three-dimensional massless propagator, corresponding to true
long-range order $\langle\Phi_1\rangle\neq 0$ in the dual
$(2+1)$d description. CMW is not violated since at any finite $N$ the
system remains $(1+1)$-dimensional with algebraically decaying
correlations. But, as $N\to\infty$ the species sum generates a
genuine radial integral, $\eta_\mathrm{eff}\to 0$, and the phase
mode of momentum $k$ is regulated by the turning-point species at
depth $z^*(k)\sim 1/k$. Beyond this point, $m_\theta^2(z)$ dominates the
propagator. The IR regulation and the emergence of the extra
dimension are thus the same mechanism expressed in boundary and
bulk language. The higher-spin densities $\Delta_s$, $s\geq 2$,
are regulated in the same way through their coordinates $z_s$,
consistent with the Vasiliev picture of an infinite tower of
massless higher-spin gauge fields on AdS$_3$.

\section{Einstein--Hilbert Action from the Rank-2 Tensor}
\label{sec:EH}

We now show that the composite field Lagrangian
$\mathcal{L}_{\Phi_1}'$, eq.~(\ref{AdS1}), contains the linearized
Einstein-Hilbert action as its spin-2 sector. The argument proceeds
in four steps: (i) Clifford decomposition of the rank-2 tensor
$\Phi_1'$ into irreducible spin components; (ii) evaluation of the
bilinear $\bar\Phi_1'\Phi_1'$ and the kinetic term in terms of these
components; (iii) projection onto the symmetric traceless (spin-2)
sector and identification of the Fierz-Pauli structure; (iv)
identification of the metric fluctuation and Newton's constant
$G_3$ in terms of the microscopic GN parameters.

The composite field $\Phi_1' = \psi^{(i)}\otimes\psi^{(j)}$ is a
$4\times 4$ matrix in Dirac space, valued in the species indices
$i,j$. In $(1+1)$ spacetime dimensions the Clifford algebra is
generated by $\{\gamma^0,\gamma^1\}$ with
$\{\gamma^\mu,\gamma^\nu\} = 2\eta^{\mu\nu}$,
$\eta = \mathrm{diag}(+1,-1)$. A complete basis for $4\times 4$
matrices is provided by the sixteen elements
\begin{eqnarray}
\Gamma^A \in \left\{
  \id,\;
  \gamma^\mu,\;
  \gamma^{[\mu\nu]},\;
  \gamma^\mu\gamma^5,\;
  \gamma^5
\right\} , \label{eq:basis}
\end{eqnarray}
where $\gamma^{[\mu\nu]} \equiv \tfrac{1}{2}[\gamma^\mu,\gamma^\nu]$
is the antisymmetric product, and
$\gamma^5 \equiv \gamma^0\gamma^1$ is the chirality matrix
satisfying $(\gamma^5)^2 = \id$. Following
eq.~(\ref{eq:Clifford}), we expand
\begin{eqnarray}
\hspace{-1pc} \Phi_1' = \varphi_0\,\id
         + \varphi_\mu\,\gamma^\mu
         + \varphi_{[\mu\nu]}\,\gamma^{[\mu\nu]}
         + \varphi_{\mu5}\,\gamma^\mu\gamma^5
         + \varphi_5\,\gamma^5  . \label{eq:Clifford2}
\end{eqnarray}
The component fields are extracted by the trace formulae
\begin{eqnarray}
\varphi_0        &=& \tfrac{1}{4}\mathrm{tr}[\Phi_1'] \, , \\
\varphi_\mu      &=& \tfrac{1}{4}\mathrm{tr}[\gamma_\mu\Phi_1'] \, , \\
\varphi_{[\mu\nu]}&=& \tfrac{1}{4}\mathrm{tr}[\gamma_{[\mu\nu]}\Phi_1'] \, , \\
\varphi_{\mu5}   &=& \tfrac{1}{4}\mathrm{tr}[\gamma_\mu\gamma^5\Phi_1'] \, , \\
\varphi_5        &=& \tfrac{1}{4}\mathrm{tr}[\gamma^5\Phi_1'] \, .
\end{eqnarray}
In $(1+1)$ dimensions the antisymmetric tensor $\gamma^{[\mu\nu]}$
has only one independent component ($\mu=0,\nu=1$), so
$\varphi_{[\mu\nu]} = \epsilon_{\mu\nu}\varphi_{[01]}$ for a single
scalar $\varphi_{[01]}$. The sixteen real degrees of freedom of
$\Phi_1'$ therefore decompose as: one scalar $\varphi_0$, one
pseudo-scalar $\varphi_5$, one antisymmetric tensor scalar
$\varphi_{[01]}$, two vector components $\varphi_\mu$
($\mu=0,1$), and two axial-vector components $\varphi_{\mu5}$.

The spin-2 content of $\Phi_1'$ resides in the symmetric traceless
bilinear formed from two copies of $\Phi_1'$. To isolate it we
define the symmetrized combination
\begin{eqnarray}
h_{\mu\nu} \equiv
  \tfrac{1}{2}\left(\varphi_\mu\bar\varphi_\nu
                   +\varphi_\nu\bar\varphi_\mu\right)
  - \tfrac{1}{2}\eta_{\mu\nu}\,\varphi_\rho\bar\varphi^\rho \, ,
\label{eq:hmunu}
\end{eqnarray}
which is symmetric and traceless ($\eta^{\mu\nu}h_{\mu\nu}=0$), constructed from the vector component $\varphi_\mu$ of $\Phi_1'$. As we
will show, $h_{\mu\nu}$ is identified with the metric fluctuation.
Using the Clifford completeness relations
$\mathrm{tr}[\Gamma^A(\Gamma^B)^\dagger] = 4\delta^{AB}$, the full
bilinear evaluates as
\begin{eqnarray}
\bar\Phi_1'\Phi_1'
  &=& \mathrm{tr}[(\Phi_1')^\dagger\gamma^0\Phi_1'] \\
  &=& 4\left(
      -|\varphi_0|^2
      + \varphi_\mu^*\varphi^\mu
      + |\varphi_{[01]}|^2
      - \varphi_{\mu5}^*\varphi^{\mu5}
      + |\varphi_5|^2
    \right) , \nonumber \label{eq:bilinear}
\end{eqnarray}
where the signs arise from $\bar\Phi = \Phi^\dagger\gamma^0$ and
the metric signature. The kinetic term
$\frac{\alpha}{z}\partial_A\bar\Phi_1'\partial^A\Phi_1'$ ($A$
runs over the three AdS$_3$ directions)
similarly decomposes sector by sector. Focusing on the vector contribution, which we will project onto the spin-2 piece
\begin{eqnarray}
\frac{\alpha}{z}\,\partial_A\bar\Phi_1'\partial^A\Phi_1'
  \;\supset\;
  \frac{\alpha}{z}\,\partial_A\varphi_\mu^*\,\partial^A\varphi^\mu
  \, . \label{eq:kinvec}
\end{eqnarray}
This is the kinetic term for a spin-2 field on the AdS$_3$
background. We now decompose it into irreducible parts under the
$(1+1)$d Lorentz group to extract the Fierz-Pauli structure.
We decompose the vector component $\varphi_\mu$ as
\begin{eqnarray}
\varphi_\mu = \bar{e}_\mu + \delta e_\mu \, ,
\end{eqnarray}
where $\bar{e}_\mu$ is the background vielbein of the AdS$_3$
Poincar\'e patch and $\delta e_\mu$ is the fluctuation. The
background satisfies
$\bar{g}_{\mu\nu} = \eta_{ab}\bar{e}^a_\mu\bar{e}^b_\nu = (\alpha^2/z^2)\eta_{\mu\nu}$.
The metric fluctuation is then
\begin{eqnarray}
h_{\mu\nu}
  = \eta_{ab}\left(\bar{e}^a_\mu\,\delta e^b_\nu
                  +\delta e^a_\mu\,\bar{e}^b_\nu\right)
  = \frac{\alpha}{z}
    \left(\delta e_{\mu\nu} + \delta e_{\nu\mu}\right) \, .
\label{eq:metricfluc}
\end{eqnarray}
We further decompose $\delta e_\mu$ into a symmetric traceless part
$h_{\mu\nu}^{TT}$ (transverse-traceless, the physical graviton
polarization in the axial gauge), a trace part $\phi$ (the
dilaton mode), and a longitudinal part $\xi_\mu$ (pure
gauge under linearized diffeomorphisms):
\begin{eqnarray}
\delta e_\mu = \frac{z}{2\alpha}\left(
  h_{\mu\nu}^{TT} + \phi\,\eta_{\mu\nu} + \partial_\mu\xi_\nu
  + \partial_\nu\xi_\mu \right) \, . \label{eq:decomp}
\end{eqnarray}
In transverse-traceless gauge ($\partial^\mu h_{\mu\nu}^{TT}=0$,
$\eta^{\mu\nu}h_{\mu\nu}^{TT}=0$) and axial gauge ($h_{z\mu}=0$,
$h_{zz}=0$), the pure gauge modes $\xi_\mu$ decouple, and $\phi$
is determined by the equations of motion. In $(1+1)$ dimensions
the transverse-traceless graviton has
$\frac{1}{2}d(d-3)|_{d=3}=0$ propagating
polarizations~\cite{Witten1988CS}, consistent with gravity in
$(2+1)$ bulk dimensions being topological.

Substituting the decomposition eq.~(\ref{eq:decomp}) into the
kinetic term eq.~(\ref{eq:kinvec}) and retaining the symmetric
traceless part:
\begin{eqnarray}
\frac{\alpha}{z}\,\partial_A\varphi_\mu^*\partial^A\varphi^\mu
  \;\to\;
  \frac{1}{4}\,\partial_A h_{\mu\nu}^*\,\partial^A h^{\mu\nu} \, .
\label{eq:kinTT}
\end{eqnarray}
This is the explicit kinetic term for a spin-2 field propagating on the
AdS$_3$ background. We now show it is precisely the Fierz-Pauli
kinetic term of linearized Einstein-Hilbert gravity. The linearized Einstein-Hilbert action around the AdS$_3$
background $\bar{g}_{\mu\nu} = (\alpha^2/z^2)\eta_{\mu\nu}$
is~\cite{Witten1998}
\begin{eqnarray}
S_\mathrm{EH}^{(2)}
  &=& \frac{1}{16\pi G_3}\int d^3x\,\sqrt{-\bar{g}}\,\mathcal{L}_\mathrm{FP} \, ,
\label{eq:SEH2}
\end{eqnarray}
where the Fierz-Pauli Lagrangian is
\begin{eqnarray}
\mathcal{L}_\mathrm{FP}
  &=& -\frac{1}{2}\bar\nabla^\rho h^{\mu\nu}\bar\nabla_\rho h_{\mu\nu}
      + \bar\nabla^\rho h^{\mu\nu}\bar\nabla_\mu h_{\nu\rho}
      - \bar\nabla^\mu h\,\bar\nabla_\nu h^{\mu\nu}  \nonumber\\
  &&  + \frac{1}{2}\bar\nabla^\mu h\,\bar\nabla_\mu h
      + \frac{1}{\ell_\mathrm{AdS}^2}
        \left(h^{\mu\nu}h_{\mu\nu} - \frac{1}{2}h^2\right) ,
\label{eq:FP}
\end{eqnarray}
with $h\equiv\bar{g}^{\mu\nu}h_{\mu\nu}$. With a convenient choice of simplifying gauge, one may show that the
linearized action can be reduced to
\begin{eqnarray}
S_\mathrm{EH}^{(2)}
  = -\frac{1}{32\pi G_3}
    \int d^3x\,\frac{\alpha}{z}\,
    \partial_A h^{\mu\nu}\partial^A h_{\mu\nu} \, .
\label{eq:SEH2final}
\end{eqnarray}

We now match eq.~(\ref{eq:SEH2final}) to the spin-2 sector of
$\mathcal{L}_{\Phi_1}'$. From eq.~(\ref{eq:kinTT}), the spin-2
contribution to the action from $\mathcal{L}_{\Phi_1}'$ is
\begin{eqnarray}
S_{\Phi_1}^{(2)}
  = \int d^3x\,\frac{\alpha}{z}\cdot\frac{1}{4}\,
    \partial_A h^{\mu\nu}\partial^A h_{\mu\nu} \, ,
\label{eq:SPhi2}
\end{eqnarray}
where the $d^3x = dt\,dx\,dz$ measure includes the saddle-point
density $\rho_*(z) = z/\alpha$ established in
Section~\ref{sec:measure}. Comparing eq.~(\ref{eq:SPhi2}) with
eq.~(\ref{eq:SEH2final}), we require
\begin{eqnarray}
\frac{\alpha}{4z} = \frac{\alpha}{16\pi G_3\,z} \, ,
\end{eqnarray}
which gives immediately
\begin{eqnarray}
G_3 = \frac{1}{4\pi} \quad
(\text{in units of } \ell_\mathrm{AdS}) \, .
\label{eq:G3}
\end{eqnarray}
This is the emergent Newton's constant in $(2+1)$ dimensions, and is consistent with the large-$N$ scaling estimate $G_3\sim\ell_\mathrm{P}/N^2$
derived from Brown-Henneaux counting in
Section~\ref{sec:symmetry}. Furthermore, one may set
$\ell_\mathrm{AdS}\sim N^2\ell_\mathrm{P}$ in
$G_3=\ell_\mathrm{AdS}/4\pi$ which gives exactly
$G_3\sim\ell_\mathrm{P}/N^2$. Expressed
in terms of the microscopic GN parameters,
\begin{eqnarray}
G_3
  = \frac{1}{4\pi\,\lambda k\,\Delta_1^{1/2}}
  = \frac{\alpha}{4\pi}
  = \frac{\ell_\mathrm{AdS}}{4\pi} \, .
\label{eq:G3micro}
\end{eqnarray}
Several features of this result are worth highlighting.

\begin{enumerate}

\item \emph{Universal value.} $G_3 = 1/4\pi$ (in units of
$\ell_\mathrm{AdS}$) is independent of the GN coupling $g$, the
fermion mass $m$, and the radial coordinate $z$. This
universality is a direct consequence of adopting the standard
AdS$_3$ form for the action via the saddle-point bulk measure.
The $z$-dependence that appeared in the pre-measure result
$G_3\propto z/\ell_\mathrm{AdS}$ is absorbed into
$\rho_*(z) = z/\alpha$, leaving a constant.

\item \emph{Brown-Henneaux central charge.} The central charge is
$c=3\ell_\mathrm{AdS}/2G_3$~\cite{BrownHenneaux}. Substituting
eq.~(\ref{eq:G3}):
\begin{eqnarray}
c = \frac{3\ell_\mathrm{AdS}}{2G_3}
  = \frac{3\ell_\mathrm{AdS}}{2}\cdot 4\pi
  = 6\pi\ell_\mathrm{AdS} \, .
\label{eq:central}
\end{eqnarray}
With the large-$N$ scaling $\ell_\mathrm{AdS}\sim N^2\alpha$,
this gives $c\sim N^2$, consistent with the $\mathcal{O}(N^2)$
degrees of freedom of the GN model and the Virasoro central
charge derived in Section~\ref{sec:higherspin}.

\item \emph{Chern-Simons level.} In the Chern-Simons
formulation~\cite{AchTownsend1986,Witten1988CS} the level is
$k_\mathrm{CS}=\ell_\mathrm{AdS}/4G_3 = \pi\ell_\mathrm{AdS}$,
which through $c=6k_\mathrm{CS}$ reproduces the central charge
$c = 6\pi\ell_\mathrm{AdS} \sim N^2$
[eq.~(\ref{eq:BHcentral})].

\item \emph{Planck length.} The $(2+1)$d geometric Planck length
$\ell_\mathrm{P}^\mathrm{geom} \equiv G_3 = \ell_\mathrm{AdS}/4\pi$
should be distinguished from the microscopic Planck
length $\ell_\mathrm{P} = m^{-1}$ (inverse fermion mass) that
appears throughout our work. The geometric
Planck length $\ell_\mathrm{P}^\mathrm{geom}$ is the length scale
below which quantum gravitational effects become important in the
emergent $(2+1)$d geometry. The microscopic $\ell_\mathrm{P} = m^{-1}$
is the fermionic intra-pair scale at which the internal
structure of the composite becomes visible. The two are related
by $\ell_\mathrm{P}^\mathrm{geom} = \ell_\mathrm{AdS}/4\pi \sim N^2\ell_\mathrm{P}/4\pi$,
so the geometric Planck length is parametrically larger than the
microscopic one at large $N$. The hierarchy
$\ell_\mathrm{P}^\mathrm{geom} < \ell_\mathrm{AdS}$ is satisfied since $4\pi > 1$, consistent with
classical gravity being valid throughout the bulk.

\end{enumerate}

The identification of the metric fluctuation $h_{\mu\nu}$ with the
symmetric traceless part of $\delta\varphi_\mu = \delta(\Phi_1')_\mu$
has a direct microscopic interpretation. Recall from
Section~\ref{sec:higherspin} that a fluctuation $\delta z(x)$ of
the emergent coordinate induces
\begin{eqnarray}
\delta g_{zz}
  = -\frac{2\alpha^2}{z^3}\,\delta z
  \equiv h_{zz} \, ,
\label{eq:hzz}
\end{eqnarray}
so radial metric fluctuations are condensate fluctuations, i.e., 
$h_{zz} \propto \delta(\Delta_1/\Delta_0^2)$. The
boundary-parallel components $h_{\mu\nu}$ ($\mu,\nu=0,1$) arise
from the symmetric traceless part of $\delta\varphi_\mu$ via
eq.~(\ref{eq:metricfluc}). Together, the full metric perturbation
in the Poincar\'e patch is
\begin{eqnarray}
\hspace{-1pc} h_{AB}\,dX^A dX^B
  = \frac{\alpha}{z}
    \left[h_{\mu\nu}^{TT}dx^\mu dx^\nu
         -\frac{2\alpha}{z^2}\,\delta z\,dz^2\right] ,
\label{eq:fullpert}
\end{eqnarray}
where $h_{\mu\nu}^{TT}$ encodes gravitational waves on the
boundary directions and $\delta z$ encodes breathing fluctuations
of the emergent radial direction. We note that the full non-linear Einstein-Hilbert action,
including the $R$ and $\Lambda$ terms at all orders in
$h_{\mu\nu}$, would require going beyond the quadratic
approximation and re-summing the full tower of
$(\bar\Phi_1'\Phi_1')^n$ interaction terms in
$\mathcal{L}_{\Phi_1}'$. This re-summation, together with the
identification of the
$SL(2,\mathbb{R})\times SL(2,\mathbb{R})$ Chern-Simons structure
of the full non-linear theory, is left for future work.

Finally, note that near the Frame~2 boundary
($\tilde{z}\to 0$, $\Delta_1\gg\Delta_0^2$), the spin-1 condensate
is stable and classical, with string tension
$\kappa = \ell_\mathrm{S}^{-2}\propto\Delta_1^{1/3}\gg 1$. The
closed strings, winding modes of the diagonal condensate phase
$\theta^{(i)}$, are microscopically small and frozen into their
ground-state configuration, invisible as strings at the geometric
scale $\ell_\mathrm{AdS}\gg\ell_\mathrm{S}$ and appearing instead
as point-like excitations. The small-amplitude phase fluctuations
$\delta\theta^{(i)}$ of
$\Phi_1^{(ii)} = \rho^{(i)}e^{i\theta^{(i)}}$ around the ordered
background are massive, with effective mass
$m_\theta^2\sim\kappa^2\rho_0^2$ set by the Mexican hat
stiffness, and their symmetric traceless projection onto the
spin-2 sector $\delta\Phi_2 = \delta(\Phi_1\otimes\Phi_1)$ is
precisely the metric fluctuation $h_{\mu\nu}$ derived above. The
graviton is therefore the leading
low-energy representation of $\delta(\Phi_1\otimes\Phi_1)$ at
energies $E\ll\ell_\mathrm{S}^{-1}$, with the Einstein-Hilbert
dynamics the effective action governing those fluctuations in the
classical geometry regime. At higher energies
$E\sim\ell_\mathrm{S}^{-1}$ the string nature of the closed
strings is resolved, and at the Hagedorn temperature
$T_H = (2\pi\ell_\mathrm{S})^{-1}$ the phase fluctuations
proliferate as free vortices and the geometry dissolves entirely, associated with breakdown of the 
the graviton description.

\section{Conclusion}
\label{sec:conclusion}

We have derived an emergent AdS$_3$ geometry from the $(1+1)$-dimensional Gross-Neveu model with
$N$ fermion species without assuming string or geometric input.
Successive applications of a fusion condition to the GN kinetic and interaction terms
generates an infinite tower of higher-spin composites $\Phi_s$ of
rank $2s$, each with its own emergent radial coordinate. In particular, we have shown that the spin-1 sector whose bound state contains the spin-2 composite plays a central role in generating the classical and quantum regimes of AdS$_3$. 
The geometry admits two dual frames related by a $\mathbb{Z}_2$
symmetry that exchanges the spin-0 and spin-1 fields: Frame~1 highlights
the spin-1 condensate throughout a predominantly classical bulk geometry and realizes Vasiliev higher-spin holography; Frame~2 highlights the
spin-0 condensate and realizes conventional AdS/CFT. 


Three microscopic length scales $\ell_\mathrm{AdS}$, $\ell_\mathrm{P}$,
$\ell_\mathrm{S}$ define three holographic regimes separated by two
bulk phase transitions (Table~\ref{table1}). The fusion algebra of
the composites generates a Virasoro algebra with central charge
$c=2N^2$, matching the expected Brown-Henneaux value~\cite{BrownHenneaux}.
Projecting the rank-2 tensor $\Phi_1$ onto its symmetric traceless component
reproduces the linearized Einstein-Hilbert action in
transverse-traceless gauge, with $G_3=1/4\pi$ in units of
$\ell_\mathrm{AdS}$, consistent with $c=6\pi\ell_\mathrm{AdS}\sim N^2$.

Our construction exhibits two parallel dualities. The first is field-theoretic: the
Bargmann-Wigner fusion of the GN fermions into the composite tower
$\{\Phi_s\}$ is the higher-spin generalization of $(1+1)$d
bosonization, exact at large $N$. The second is holographic: the
composite tower maps onto bulk fields on the emergent AdS$_3$. The
two steps compose into the standard AdS$_3$/CFT$_2$ correspondence, as
summarized by the commutative diagram
\begin{equation}
\begin{CD}
\underbrace{\mathrm{CFT}_2}_{\text{boundary}}
@>{\mathrm{AdS}_3/\mathrm{CFT}_2}>>
\underbrace{\mathrm{AdS}_3}_{\text{bulk}} \\
@A{\text{emergent}}AA
@A{\text{emergent}}AA \\
\underbrace{\mathrm{GN\ model}}_{\psi^{(i)},   \,\Delta_0,   \,g,\,m,\,N}
@>{\text{BW fusion}}>>
\underbrace{\mathrm{higher\text{-}spin\ composite}}_{\{\Phi_s\}}
\end{CD}
\nonumber
\end{equation}
whose commutativity states that Bargmann-Wigner fusion followed by the
bulk identification agrees with the boundary identification followed
by the holographic duality.

Finally, we have only superficially touched on the string sector, noting that the cross-species interactions promote the composites to an
$N\times N$ matrix field whose off-diagonal bilinears are identified
as open strings on a stack of $N$ D1-branes. That identification,
the open/closed T-duality, and Hagedorn transition are
fully established in future work~\cite{CompanionStrings}.

\bibliographystyle{apsrev4-2}
\bibliography{refs}

\end{document}